\documentclass[sigconf]{acmart}

\usepackage{booktabs} % For formal tables

\acmConference[KDD'25]{The 30th ACM SIGKDD International Conference on Knowledge Discovery and Data Mining}{August X - Y, 2025}{Toronto, Canada}
\usepackage{amscd,amsfonts,amsbsy,rotating}
\usepackage{balance}
\usepackage{graphicx}
\usepackage{epsfig,epstopdf}
\usepackage{subfigure}
\usepackage{multirow}
\usepackage{booktabs}
\usepackage{color,xcolor}
\usepackage{url}
\usepackage{latexsym,bm}
\usepackage{enumitem,balance,mathtools}
\usepackage{wrapfig}
\usepackage{euscript}
\usepackage{algorithm}
\usepackage{algorithmic}
\usepackage{ifpdf}
\usepackage{diagbox}
\usepackage{caption}
\usepackage{makecell}
\usepackage{subfigure}
\usepackage[leftcaption]{sidecap}
\usepackage{textcomp}
\usepackage{booktabs}
\renewcommand{\thefootnote}{\fnsymbol{footnote}}

\DeclareMathOperator*{\argmax}{arg\,max}

\usepackage{array}
\newcolumntype{N}{@{}m{0pt}@{}}

\usepackage{lipsum}

\newcommand{\minisection}[1]{\vspace{5pt}\noindent\textbf{#1}}

\begin{document}
    \title{Why Not Together? A Multiple-Round Recommender System\\ for Queries and Items}
	\author{Jiarui Jin$^{1,*}$, Xianyu Chen$^{1}$, Weinan Zhang$^1$, Yong Yu$^1$, Jun Wang$^2$}
	\affiliation{
$^1$Shanghai Jiao Tong University, $^2$University College London
}
\email{{jinjiarui97, xianyujun, wnzhang, yyu}@sjtu.edu.cn, jun.wang@cs.ucl.ac.uk}
\renewcommand{\shortauthors}{Jiarui Jin et al.}
	\renewcommand{\shorttitle}{MAGUS}
	
	\settopmatter{printacmref=false}
	
    \begin{abstract}
    A fundamental technique of recommender systems involves modeling user preferences, where queries and items are widely used as symbolic representations of user interests.
    Queries delineate user needs at an abstract level, providing a high-level description, whereas items operate on a more specific and concrete level, representing the granular facets of user preference.
    % Query recommendations aim to forecast the queries that users are inclined to search, whereas item recommendations focus on predicting the items that users are likely to prefer.
    While practical, both query and item recommendations encounter the challenge of sparse user feedback.
    % Our premise is that both query and item recommendations share the common goal of capturing user interests, with the main distinction lying in the level of granularity at which they operate.
    % differing primarily in the level of granularity at which they operate.
    % Query recommendations primarily operate at an abstract level by anticipating user search queries, whereas item recommendations function on a more specific level by predicting user preferences for individual items.
    To this end, we propose a novel approach named \underline{\texttt{M}}ultiple-round \underline{\texttt{A}}uto \underline{\texttt{G}}uess-and-\underline{\texttt{U}}pdate \underline{\texttt{S}}ystem (\textsf{MAGUS}) that capitalizes on the synergies between both types, allowing us to leverage both query and item information to form user interests. 
    This integrated system introduces a recursive framework that could be applied to \emph{any} recommendation method to exploit queries and items in historical interactions and to provide recommendations for both queries and items in each interaction round. 
    % This integrated system, denoted as \underline{\texttt{M}}ultiple-round \underline{\texttt{A}}uto \underline{\texttt{G}}uess-and-\underline{\texttt{U}}pdate \underline{\texttt{S}}ystem (\textsf{MAGUS}), introduces a recursive framework that could be applied to \emph{any} recommendation method.
    % \textsf{MAGUS} enables the utilization of both queries and items in historical interactions and provides recommendations for both queries and items at each interaction round.
    Concretely, \textsf{MAGUS} first represents queries and items through combinations of categorical words, and then constructs a relational graph to capture the interconnections and dependencies among these individual words and word combinations.
    % Subsequently, we construct a relational graph to capture the interconnections and dependencies among these individual words and word combinations.
    In response to each user request, \textsf{MAGUS} employs an offline tuned recommendation model to assign estimated scores to words representing items; and these scores are subsequently disseminated throughout the graph, impacting each individual word or combination of words.
    Through multiple-round interactions, \textsf{MAGUS} initially guesses user interests by formulating meaningful word combinations and presenting them as potential queries or items.
    Subsequently, \textsf{MAGUS} is updated based on user feedback, enhancing its recommendations iteratively.
    % and then it is updated according to user feedback. 
    % In multiple-round interactions, \textsf{MAGUS} endeavors to initially deduce user interests by formulating meaningful word combinations, presenting them as potential queries or items.
    % As the interaction progresses, \textsf{MAGUS} receives user feedback, which triggers the application of a newly designed label propagation algorithm.
    % In multiple-round interactions, \textsf{MAGUS} endeavors to initially deduce user interests by formulating meaningful word combinations, presenting them as potential queries or items.
    % As the interaction progresses, \textsf{MAGUS} receives user feedback, which triggers the application of a newly designed label propagation algorithm.
    % This proposed algorithm efficiently models the influence of feedback across interconnected nodes with the graph, leading to the adjustment of prediction scores.
    Empirical results from testing 12 different recommendation methods demonstrate that integrating queries into item recommendations via \textsf{MAGUS} significantly enhances the efficiency, with which users can identify their preferred items during multiple-round interactions.
    % Empirical results with 12 different recommendation methods demonstrate that integrating queries into item recommendations through \textsf{MAGUS} yields significant advantages for users in terms of efficiently identifying their preferred items in multiple-round interactions recommendations.
    % A notable application scenario for \textsf{MAGUS} is exploratory search, where users lack a precise picture of item preference.
    % \textsf{MAGUS} offers users the capability to navigate and compare items prior to refining their query expressions or opting for alternative queries.
    % that better align with their evolving preferences.
    \end{abstract}
	
	\settopmatter{printacmref=false}
	
\maketitle

% {
\renewcommand{\thefootnote}{\fnsymbol{footnote}}
\footnotetext[1]{Work done during Jiarui Jin's visit at University College London.}
% }

\section{Introduction}
\begin{figure}[t]
	\centering
 \includegraphics[width=1.00\linewidth]{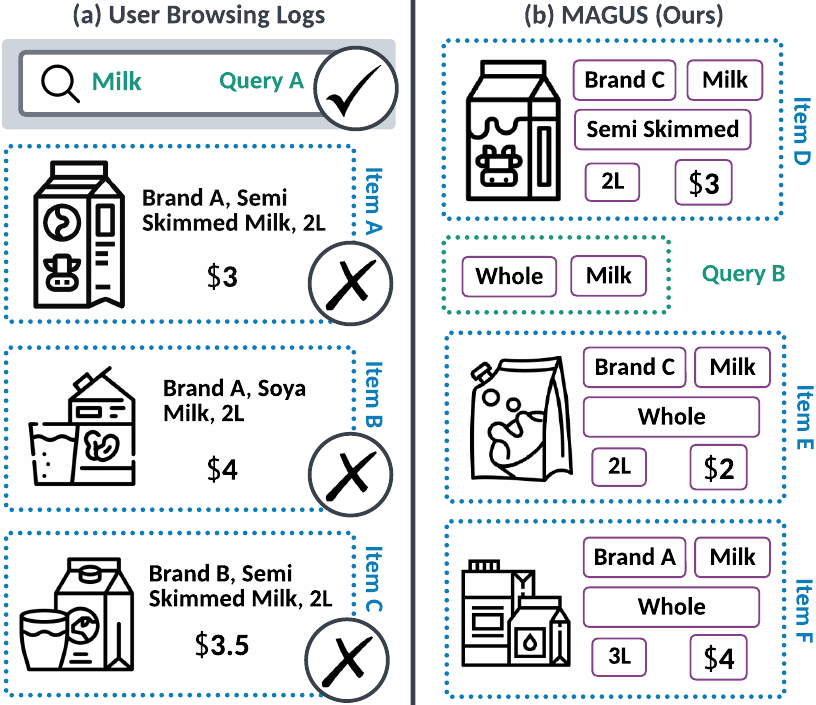}
	\vspace{-6mm}
	\caption{An illustrated example of \textsf{MAGUS} recommending both queries (e.g., query B) and items (e.g., items D, E, and F), as shown in (b). 
    This framework effectively leverages both queries (e.g., query A) and items (e.g., items A, B, and C) present in the browsing logs, serving to alleviate the data sparsity issue of user feedback, as shown in (a).
	}
	\label{fig:motivation}
	\vspace{-4mm}
\end{figure}

Information retrieval with applications in search engines \citep{liu2009learning} and recommender systems \citep{resnick1997recommender} aims to offer preferred or favored items to users.
A fundamental technique is the capture of user interest and demands, predominantly characterized by queries and items.
Queries concentrate on the prediction of user-search queries at an abstract level, whereas items operate on individual items at a more specific level. 
Existing search engine algorithms (commonly known as learning-to-rank algorithms) and recommendation methods always utilize query information and user profiles as inputs and employ user-browsed items as supervision signals.
However, we argue that queries, being capable of describing user needs, should also be employed as a form of supervision.
Namely, relying exclusively on items as the sole basis for supervision could result in suboptimal solutions.
Taking Figure~\ref{fig:motivation}(a) as an example, a user uses query A \texttt{Milk}, prompting the recommender system to return items A, B, and C; and all these items receive negative feedback.
Previous recommendation methods, centered solely on items, would refrain from suggesting additional milk items to the user.
However, these approaches overlook the explicit indication of the user's interest in milk through query A.
This particular case motivates us to incorporate query information within item recommendations, given that both elements fundamentally aim to describe user interests.

In this paper, we propose a novel  \underline{\texttt{M}}ultiple-round \underline{\texttt{A}}uto \underline{\texttt{G}}uess-and-\underline{\texttt{U}}pdate \underline{\texttt{S}}ystem (\textsf{MAGUS}), a multiple-round recommendation framework applicable to \emph{any} recommendation method.
\textsf{MAGUS} allows a recommender system to leverage both queries and items in historical interactions to offer recommendations for queries and items at each round, thus addressing the data sparsity issue inherent in user feedback. 
In the design of \textsf{MAGUS}, we are, at least, required to handle the following challenges.  

\begin{itemize}[topsep = 3pt,leftmargin =5pt]
\item \textbf{[C1]}
How to draw connections between queries and items?
Taking Figure~\ref{fig:motivation} as an example, a core challenge of jointly considering queries and items lies in creating a unified metric for evaluating queries and items (e.g., query B and items D, E, and F).
% item recommendations primarily involve comparing various milk items, whereas, query recommendations aim to suggest queries relevant to milk. 
% Hence, the fundamental challenge lies in creating a unified metric for evaluating both queries and items simultaneously.
  
\item \textbf{[C2]}
How to model interdependence among queries? 
Unlike items, queries exhibit a significant degree of dependence.
Here are three possible scenarios for each query-query pair:
(i) Mutual improvement: selecting one query increases the likelihood of selecting the other query in the following round (e.g., selecting \texttt{Milk} would raise the probability of selecting \texttt{Whole} \texttt{Milk} in the next round).
(ii) Mutual inhibition: selecting one query decreases the probability of selecting the other query in the following round (e.g., if a user selects \texttt{Milk}, it is unlikely that she would select \texttt{Beef} because milk and beef belong to distinct categories.
(iii) Mutual Independence: the selection of one query has minimal or negligible effects on the user's decision regarding the other query (e.g., selecting \texttt{Milk} does not significantly influence the user's preference for \texttt{On} \texttt{Sale}).
\item \textbf{[C3]}
How to efficiently make use of user feedback at each round within strict online latency requirements?
One of the primary advantages of multiple-round recommendations lies in its ability to progressively approach user interests through iterative guesses and updates.
However, this recursive framework presents a challenge in terms of effectively leveraging user feedback while maintaining strict online latency requirements.
\end{itemize}

We recognize that both queries and items act as representations of user interests, albeit at different levels of granularity.
In this context, both queries and items can be regarded as combinations of specific categorical words describing user interests.
As exemplified in Figure~\ref{fig:motivation}(b), item A could be expressed as the combination of words \texttt{Brand} \texttt{A}, \texttt{Milk}, \texttt{Semi} \texttt{Skimmed}, \texttt{2L}, and \texttt{\$3}; and query B can be organized as the combination of words \texttt{Whole} and \texttt{Milk}.
Therefore, queries and items could be connected through the shared use of words, thereby addressing [C1]. 
To capture the dependence among these words, we construct a relational graph, where the nodes correspond to individual words or combinations of words, and the edges represent three types of dependence introduced in [C2].
We illustrate the graph in Figure~\ref{fig:overview}(a).

% To address the aforementioned challenges, it's important to recognize that both queries and items act as representations of user interests, albeit at different levels of granularity.
% Namely, queries capture user needs at a higher and more abstract level, while items provide specific and detailed products that satisfy those needs.
% In this context, both queries and items can be regarded as combinations of specific categorical words.

In each user session, \textsf{MAGUS} starts with an arbitrary offline tuned recommendation model to assign an initial prediction score to every node representing an item in the graph, e.g., items A, B, and C in Figure~\ref{fig:overview}(a).
Subsequently, \textsf{MAGUS} propagates these scores to associated words throughout the graph, as demonstrated in Figure~\ref{fig:overview}(b).
During each interaction round, \textsf{MAGUS} selects those nodes with the highest scores.
These nodes would either comprise artificial queries or specific items.

In the above multiple-round setting, \textsf{MAGUS} is designed to receive user feedback during each interaction round.
To effectively harness this feedback while adhering to stringent online latency requirements (i.e., [C3]), \textsf{MAGUS} employs a label propagation algorithm to simulate the influence of user feedback on the relevant words, as depicted in Figure~\ref{fig:overview}(c).
In our approach, we also introduce a feature propagation method to determine edge weights within the graph.
These weights enable us to perform weighted propagation as illustrated in Figure~\ref{fig:overview}(b) and (c). 

% In each session, we start with an offline-tuned recommender system to assign an initial predicted score for every node representing an item in the graph, e.g., \texttt{Item} \texttt{A}, \texttt{Item} \texttt{B}, \texttt{Item} \texttt{C} in Figure~\ref{fig:overview}(a).
% Subsequently, we propagate these scores to associated words throughout the graph, as demonstrated in Figure~\ref{fig:overview}(b).
% During each round, \textsf{MAGUS} conducts a sorting operation on all the nodes and generates a list comprising nodes with high scores. 
% These high-scored nodes would represent either specific items or artificial queries.
% We also show that Large Language Models (LLMs) can decorate these fabricated queries more closely with real queries.

Given that our principal aim is to identify an item aligned with user preferences, we have formulated an innovative multiple-round recommendation simulator for evaluation purposes. 
Within this simulation environment, the \textsf{MAGUS} agent operates iteratively, providing suggestions for both queries and items in each round, while the user agent responds accordingly. 
This iterative cycle continues until an item that sufficiently aligns with user requirements is discovered. Furthermore, we demonstrate that incorporating large language models into the simulator can enhance its capabilities, transforming it into a conversational recommender system.

Our extensive experiments, conducted on 3 real-world datasets and incorporating 12 diverse recommendation methods, consistently validate that \textsf{MAGUS} significantly enhances recommendation performance by effectively identifying items that align with user preferences.
Additionally, our results also demonstrate that \textsf{MAGUS} can effectively handle cases where users do not have a clear picture of what they like.

% Additionally, we also provide a particular use case of \textsf{MAGUS} called exploratory search, catering to situations where users possess vague ideas about the items they seek. 
% In such instances, \textsf{MAGUS} has the capability to present a curated selection of recommended milk items along with queries, enabling users to navigate and compare items before refining their search queries.

\section{Preliminaries}
\subsection{Problem Formulation}
\label{sec:problem}
We begin by introducing the notations used in this paper.
$\mathcal{U}$ denotes the set of users, $\mathcal{V}$ denotes the set of items, and $\mathcal{Q}$ denotes the set of queries.
For convenience, we use $\psi_\mathtt{RE}(\cdot)$, a mapping function $\psi_\mathtt{RE}:\mathcal{U}\times\mathcal{V}\rightarrow \mathbb{R}$, to denote any given offline tuned recommendation method that assigns a predicted relevance score to each item.
We denote the score of each item $v\in\mathcal{V}$ as $\psi_\mathtt{RE}(v)$\footnote{We use omit $u$ in notations (i.e., using $\psi_\cdot(v)$ instead of $\psi_\cdot(u,v)$) for simplicity.}.
Our goal is to build a joint mapping function $\psi_\mathtt{MAGUS}:\mathcal{U}\times\mathcal{A}\rightarrow \mathbb{R}$, where $\mathcal{A}=\mathcal{V}\cup\mathcal{Q}$ is the action space.
In other words, our \textsf{MAGUS}, i.e., $\psi_\mathtt{MAGUS}(\cdot)$, is capable of exploiting both query information and item information to learn to assign scores to both queries and items.

We highlight that the primary objective of \textsf{MAGUS} is to identify an item that meets the user's requirement within a session, where \textsf{MAGUS} is allowed to do multiple-round recommendations of either queries or items within each user session.
For clarity, we present the description in the context of the top-1 recommendation and discuss its extension to the top-N recommendation in Appendix~\ref{sec:analyze}.
\begin{definition}
[\emph{\textbf{\textsf{Multiple-round Recommender Systems Supporting both Queries and Items}}}]
\label{def:task}
Given a tuple $(\mathcal{U},\mathcal{V},\mathcal{Q}$, $K_\mathtt{MAX})$, in each user session (with user $u\in\mathcal{U}$), the objective of $\psi_\mathtt{MAGUS}(\cdot)$ is to recommend an item satisfying $u$'s needs within $K_\mathtt{MAX}$ rounds.
The action space of $\psi_\mathtt{MAGUS}(\cdot)$ covers both the item space and the query space, namely, $\mathcal{A}=\mathcal{V}\cup\mathcal{Q}$.
For each $k$-th round (where $k=1,\ldots, K_\mathtt{MAX}$), $\psi_\mathtt{MAGUS}(\cdot)$ recommends either an item or a query, and the user should provide either positive or negative feedback.
Formally, let $\mathcal{V}_\mathtt{TARGET}$ denote the set of items that can fulfill the user's requirements, and $a_k\in\mathcal{A}$ signifies the recommendation made by $\psi_\mathtt{MAGUS}(\cdot)$ at the $k$-th round. 
Our objective is:
\begin{equation}
\min K, \text{ s.t., }a_K\in\mathcal{V}_\mathtt{TARGET},
\end{equation}
where $K=1,\ldots,K_\mathtt{MAX}$.
In other words, an oracle $\psi_\mathtt{MAGUS}(\cdot)$ is supposed to find an item in $\mathcal{V}_\mathtt{TARGET}$ with a minimal number of rounds.
For this purpose, \textsf{MAGUS} is supplied with the historical interactions of users, including $\mathcal{H}^q_u$ representing the set of searched queries associated with each user $u$, $\mathcal{H}^+_u$ representing the set of items with positive feedback (e.g., observed and clicked) of each user $u$, and $\mathcal{H}^-_u$ representing the set of items with negative feedback (e.g., observed and not clicked) of each user $u$.
We also note that \textsf{MAGUS} can make use of $\mathcal{H}^q_u$, but $\mathcal{H}^q_u$ is not compulsory for applying \textsf{MAGUS}.
\end{definition}

\subsection{Comparisons to Previous Work}
Designing multiple-round recommender systems to identify an item meeting user requirements has been explored in the domain of conversational recommendations \citep{gao2021advances,jannach2021survey}. 
One popular direction is to combine the recommender module and the conversational module from a systematic perspective \citep{lei2020estimation,lei2020interactive,sun2018conversational,zhang2020conversational}.
For example, recent papers \citep{lei2020estimation,li2018towards} introduces a reinforcement learning framework that enables the optimizations over multiple-round interactions.
However, reinforcement learning-based approaches inherently face challenges related to the insufficient usage of labeled data and high complexity costs of deployment.
On the contrary, \textsf{ MAGUS} could be seamlessly integrated with any recommendation method to enable it to effectively exploit queries and items in historical interactions and recommend both queries and items.
Furthermore, it is essential to highlight that our recommending queries differ significantly from recommending attributes in conversational recommender systems like \citep{lei2020estimation,bi2019conversational,sun2018conversational,zhang2020conversational}.
The key distinction lies in the substantial dependence among queries such as \texttt{apple} and \texttt{macbook}, \texttt{macbook} and \texttt{macbook} \texttt{pro}, which is not as prominent among attributes such as brand and color.
% This discrepancy stems from the nature of the information being conveyed:
% Users typically refine their searches or articulate their intentions by adding or adjusting queries, while attributes are used to describe various facets or traits of an item.
In addition, in contrast to our MAGUS system, most conversational recommendation methods \citep{lei2020estimation,sun2018conversational} use reinforcement learning techniques, whereas reinforcement learning based methods heavily rely on meticulous reward function design. In comparison, the \textsf{MAGUS} system can be easily deployed into an arbitrary recommendation method, as our framework applies an offline-tuned recommendation model in a plug-and-play manner.

Our \textsf{MAGUS} is also connected to query recommendations (commonly referred to as query suggestions) \citep{ooi2015survey}, aiming at improving user search experience by providing suggestions that attempt to guess user intentions based on their past behaviors.
However, our ultimate objective differs, as we aim to identify an item rather than generate a query.

\begin{figure*}[t]
	\centering
	\includegraphics[width=1.00\linewidth]{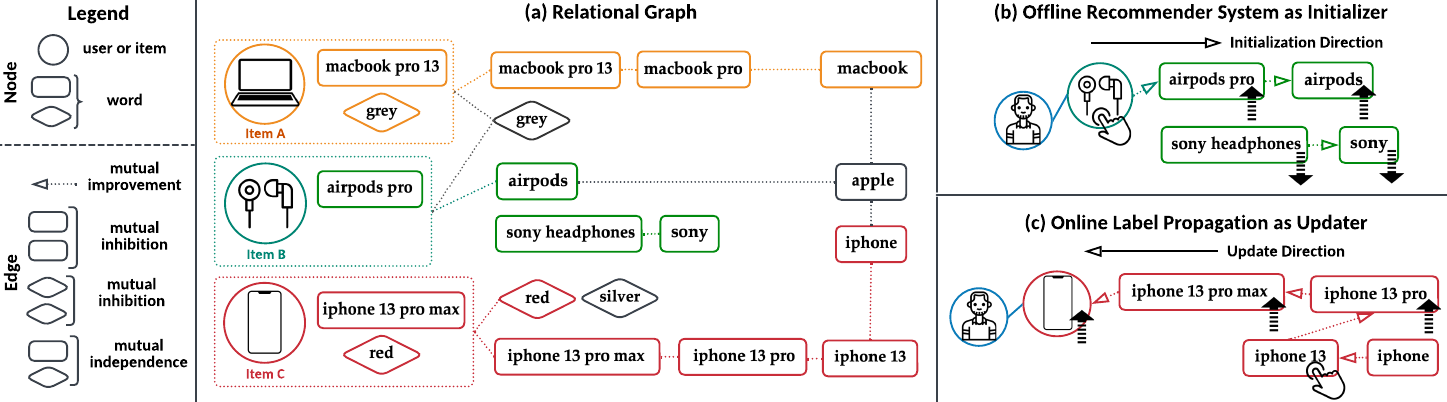}
	\vspace{-6mm}
	\caption{An illustrated example of our relational graph organizing individual words and certain combinations of words (as shown in (a)).
    To maintain clarity, we have opted not to explicitly illustrate the edges representing mutual inhibition and mutual independence.
    In each user session, 
    \textsf{MAGUS} collaborates with an offline-tuned recommendation method to initialize the node scores, where the propagation direction is from the nodes representing the items towards the other nodes (as shown in (b)).
    Subsequently, when a user provides her preference on a recommendation, \textsf{MAGUS} updates the relevant nodes by a label propagation algorithm whose propagation direction now operates from the recommended node towards the nodes representing items (as shown in (c)).   
	}
	\label{fig:overview}
	\vspace{-2mm}
\end{figure*}

\section{The MAGUS System}
\label{sec:model}
Our main idea is to recognize that queries and items both represent user interests, but they do so at varying levels of granularity.
For this purpose, we initially construct a relational graph offline to bridge queries and items (Section~\ref{sec:bridge}).
In each user session, we employ an offline-tuned recommendation model as an initializer to assign an initial score to each node (Section~\ref{sec:recommendation}).
Subsequently, we design a label propagation algorithm to iteratively update these scores throughout the session (Section~\ref{sec:update}).
Additionally, we propose a feature propagation algorithm to learn the edge weights of the graph (Section~\ref{sec:weight}), although its use is not mandatory in practice.

% we first offlinely build a relational graph to bridge queries and items 
% Then, during each user session, we employ an arbitrary offline tuned recommendation method as our  to assign  and design a label propagation algorithm to update these scores in each round  within the session. 
% In addition, we also provide a feature propagation algorithm to learn the edge weights of the graph  which is not compulsory in practice.

\subsection{Bridging Queries and Items via Graph}
\label{sec:bridge}
To establish a connection between queries and items, we use ``words'' as the bridge to build a relational graph as follows.
% A core challenge of bridging query and item recommendations is to establish a connection between queries and items.
% Here, we use ``words'' as the bridge facilitating a more seamless integration of the two recommendation types.
\begin{definition}
[\emph{\textbf{\textsf{Words and Relational Graph}}}]
\label{def:graph}
Given a tuple $(\mathcal{V},\mathcal{Q})$\footnote{A summary table of key notations is provided in the supplementary material.} 
where $\mathcal{V}$ is the set of items and $\mathcal{Q}$ is the set of queries, \textsf{MAGUS} establishes a set denoted as $\mathcal{W}$, comprised of categorical words, to serve as a connecting bridge between $\mathcal{V}$ and $\mathcal{Q}$.
To elucidate this process, \textsf{MAGUS} commerces by extracting lexical elements from the categorical attributes (a.k.a., features) of items in $\mathcal{V}$, e.g., this entails the extraction of terms \texttt{Brand} \texttt{A}, \texttt{Milk}, \texttt{Semi} \texttt{Skimmed}, \texttt{2L}, and \texttt{\$3} from item A in Figure~\ref{fig:motivation}(b).
This collection of extracted terms constitutes the set $\mathcal{W}$.
Subsequently, \textsf{MAGUS} proceeds to decompose the queries within $\mathcal{Q}$ by matching them with the words present in $\mathcal{W}$; e.g.,
query B in Figure~\ref{fig:motivation}(b) can be deconstructed into a combination of the terms \texttt{Whole} and \texttt{Milk} based on this matching process.

As a result, \textsf{MAGUS} can construct a relational graph denoted as $\mathcal{G}$ $=(\mathcal{E}, \mathcal{R})$, where $\mathcal{E}$ signifies the set of nodes and $\mathcal{R}$ denotes the set of edges.
$\mathcal{E}$ encompasses not only individual words from the vocabulary $\mathcal{W}$ but also the combinations of words that either form queries in $\mathcal{Q}$ or items in $\mathcal{V}$.
For example, as illustrated in Figure~\ref{fig:overview}(a), node (\texttt{macbook} \texttt{pro} \texttt{13} \texttt{grey}) represents item A, and node (\texttt{macbook} \texttt{pro} \texttt{13}) and node (\texttt{macbook} \texttt{pro}) are the combinations of words used to organize item A, and node (\texttt{grey}) and node (\texttt{macbook}, \texttt{apple}) are individual nodes. 
$\mathcal{R}$ contains three categories of relationships between nodes, i.e., mutual improvement (denoted as $\mathcal{R}^+$), mutual inhibition (denoted as $\mathcal{R}^-$), and mutual independence (denoted as $\mathcal{R}^{\bot}$).
In other words, $\mathcal{R}=\mathcal{R}^+\cup \mathcal{R}^- \cup \mathcal{R}^{\bot}$.
$\mathcal{R}^+$ delineates connections between pairs of nodes in situations where one node is subsumed within or constitutes a component of the other node.
For instance, node (\texttt{macbook}) is a part of node (\texttt{macbook} \texttt{pro}).
In contrast, $\mathcal{R}^-$ defines relationships between pairs of nodes that pertain to distinct items, exemplified by node (\texttt{macbook} \texttt{pro}) and node (\texttt{iphone} \texttt{pro}) associated with item A and item C respectively.
Lastly, $\mathcal{R}^\bot$ connects all the other pairs of nodes, namely, it establishes connections between pairs of nodes that can be employed to represent the same item but are characterized by a lack of shared elements.  
For example, node (\texttt{macbook} \texttt{pro}) and node (\texttt{gray}) together form item A but do not exhibit any overlapping components.
\end{definition}

Based on \emph{Definition}~\ref{def:graph}, all queries and items can be successfully mapped to their respective nodes within the graph, confirming that $\mathcal{Q}\subseteq\mathcal{E}$ and $\mathcal{V}\subseteq\mathcal{E}$.
Then, for each pair of nodes, there are three possible scenarios: 
(i) $\mathcal{R}^+$ for those where one node is a part of the other, 
(ii) $\mathcal{R}^-$ for those where one node is not a part of the other, but these nodes are compatible within a single item, 
and (iii) $\mathcal{R}^\bot$ for those where one node is not a part of the other node and these nodes are not compatible.

We note that the procedure for constructing this relational graph, as detailed earlier, does not consider all item features as ``words''; instead, we focus on utilizing key features that indicate the category of items (e.g., cat\_id, seller\_id, and brand\_id in Tmall dataset), similar to the approach in \citep{du2022learning}.
Moreover, the extraction of key item features, such as keywords, from item descriptions aligns with established tokenization practices in the field \citep{mielke2021between}, which falls outside the scope of this paper.
Our graphs, derived from user interaction data and item features, diverge from traditional knowledge graphs \citep{guo2020survey,wang2019kgat,wang2021learning} typically sourced from external databases, since our graphs are consistently accessible.

% serves as a general framework that should be customized and precisely specified according to the characteristics of the specific datasets\footnote{Take Figure~\ref{fig:overview}(a) as an example.
% In $\mathcal{R}^+$, node (\texttt{apple}) is part of node (\texttt{apple} \texttt{macbook}) where the brand \texttt{apple} is often omitted for simplicity, and in $\mathcal{R}^-$, we only draw the edges between pairs of category nodes (e.g., node (\texttt{macbook} \texttt{pro}) and node (\texttt{iphone} \texttt{13} \texttt{pro})) and pairs of color nodes (e.g., node (\texttt{gray}) and node (\texttt{red})).}.

\subsection{Recommender System as Initializer}
\label{sec:recommendation}
As described in \emph{Definition~\ref{def:task}}, for each user $u\in\mathcal{U}$, we have access to her browsed items, i.e., $\mathcal{H}^+_u$ for those with positive feedback and $\mathcal{H}^-_u$ for those with negative feedback.
Then, we can establish an offline tuned recommendation method, denoted as $\psi_\mathtt{RE}(\cdot)$, trained upon $\mathcal{H}^+_u\cup\mathcal{H}^-_u$.
$\psi_\mathtt{RE}(\cdot)$ can be any recommendation method, and we evaluate the impact of employing different $\psi_\mathtt{RE}(\cdot)$s in the experiment.
In every session with user $u$, with an offline tuned $\psi_\mathtt{RE}(\cdot)$, we can assign an estimated relevance score to each item $v\in\mathcal{V}$, denoted as $\psi_\mathtt{RE}(v)$.
% In every session, when considering a user $u\in\mathcal{U}$, given an offline-tuned recommendation method, denoted as $\psi_\mathtt{RE}(\cdot)$, we can assign an estimated relevance score for each item $v\in\mathcal{V}$, denoted as $\psi_\mathtt{RE}(v)$.
As outlined in \emph{Definition~\ref{def:graph}}, each item in $\mathcal{V}$ can correspond to a specific node in $\mathcal{E}$.
For convenience, let $\mathcal{E}_\mathtt{ITEM}$ denote the set of nodes representing items. 
Here, we use $\widehat{y}_v$ to represent the estimated relevance score for each node $v\in\mathcal{E}$.

Our initial step is to assign the estimated scores to the nodes within $\mathcal{E}_\mathtt{ITEM}$.
Formally, for each node $v\in\mathcal{E}$, this process can be expressed as follows.
\begin{equation}
\label{eqn:recommendation}
\begin{aligned}
\widehat{y}_v = \left\{
\begin{array}{cc}
\psi_\mathtt{RE}(v), & v\in\mathcal{E}_\mathtt{ITEM},\\
0, &  v\in \mathcal{E}/\mathcal{E}_\mathtt{ITEM}.
\end{array}
\right.
\end{aligned}
\end{equation}
For convenience, we normalize all scores to fall within the range of 0 to 1, namely $\widehat{y}_v\in[0,1]$ holds for all $v\in\mathcal{E}_\mathtt{ITEM}$.
% If we have access to the set of user-searched queries $\mathcal{H}^+_q$, we initialize the corresponding nodes by calculating $\widehat{y}_v = \mathtt{AVG}(\{\psi_\mathtt{RE}(v')|v'\in\mathcal{H}^+_u\}) - \mathtt{AVG}(\{\psi_\mathtt{RE}(v')|v'\in\mathcal{H}^-_u\})$ for each $v\in\mathcal{H}^+_q$, where $\mathtt{AVG}(\cdot)$ denotes the average operation. 

Next, we propagate these scores to all the other nodes in $\mathcal{E}$.
For this purpose, we assign a learnable weight to each edge in $\mathcal{R}$.
For clarity, let $w_{v'v''}$ denote the weight of the edge connecting node $v'$ and node $v''$.
With a particular recommendation method $\psi_\mathtt{RE}(\cdot)$, a simple definition of weights for the edges in $\mathcal{R}$ is:
\begin{equation}
\label{eqn:weight}
w_{vv'}=1 \text{ if }\langle v, v'\rangle\in\mathcal{R}^+\cup\mathcal{R}^-, \text{ and } w_{vv'}=0 \text{ otherwise.} 
\end{equation}
If we have access to the offline tuned representation vectors for all the items and users provided by the recommendation method, we can allocate and tune learnable weights for the edges (as later introduced in Section~\ref{sec:weight}).

We then initialize the remaining nodes by propagation on the graph.
As illustrated in Figure~\ref{fig:overview}(b), the propagation process initiates from the nodes in $\mathcal{E}_\mathtt{ITEM}$ and extends to the remaining ones. 
During the propagation, the update function of nodes in $\mathcal{E}/\mathcal{E}_\mathtt{ITEM}$ can be expressed as follows.
\begin{equation}
\label{eqn:initial}
\begin{aligned}
\widehat{y}_{v'} \leftarrow \left\{
\begin{array}{cc}
\widehat{y}_{v'} + w_{v'v''}\cdot \widehat{y}_{v''}, & \langle v',v''\rangle \in \overrightarrow{\mathcal{R}}^+,\\
\widehat{y}_{v'} - w_{v'v''}\cdot \widehat{y}_{v''}, & \langle v', v''\rangle \in \overrightarrow{\mathcal{R}}^-,\\
\widehat{y}_{v'}, & \langle v', v''\rangle \in\mathcal{R}^\bot\cup\overleftarrow{\mathcal{R}}^+\cup\overleftarrow{\mathcal{R}}^-,
\end{array}
\right.
\end{aligned}
\end{equation}
where $v'\in \mathcal{E}/\mathcal{E}_\mathtt{ITEM}$ and $v''\in\mathcal{E}$, and $\overrightarrow{\mathcal{R}}^+$ and $\overrightarrow{\mathcal{R}}^-$ are introduced to denote the update directions on the edges in $\mathcal{R}^+$ and $\mathcal{R}^-$ respectively.
Here, $\mathcal{R}^+=\overrightarrow{\mathcal{R}}^+\cup\overleftarrow{\mathcal{R}}^+$ and $\mathcal{R}^-=\overrightarrow{\mathcal{R}}^-\cup\overleftarrow{\mathcal{R}}^-$.

If we have access to the set of user-searched queries $\mathcal{H}^q_u$, the set of the searched queries for user $u$, as stated in \emph{Definition}~\ref{def:graph}, each query in $\mathcal{H}^q_u$ can correspond to a specific node in $\mathcal{E}$.
Then, let $\mathcal{E}_\mathtt{QUERY}$ denote the set of nodes representing the searched queries.
We update the estimated scores of those nodes by:
\begin{equation}
\label{eqn:query}
\widehat{y}_v \leftarrow \max (\min(\{\widehat{y}_{v'}|v'\in\mathcal{E}/(\mathcal{E}_\mathtt{QUERY}\cup\mathcal{E}_\mathtt{ITEM})\}, 1), \widehat{y}_v), v\in\mathcal{E}_\mathtt{QUERY},
\end{equation}
because those queries can largely represent user interests.

\subsection{Label Propagation as Updater}
\label{sec:update}
For each round in the session, we normalize all estimated relevance scores to fall within the range of 0 to 1, namely $\widehat{y}_v \in [0,1]$ holds for all $v\in\mathcal{E}$, through applying:
\begin{equation}
\label{eqn:norm}
\widehat{y}_v \leftarrow \frac{\widehat{y}_v-\min(\{\widehat{y}_{v'}|v'\in\mathcal{E}\})}{\max(\{\widehat{y}_{v'}|v'\in\mathcal{E}\})-\min(\{\widehat{y}_{v'}|v'\in\mathcal{E}\})}, v\in\mathcal{E}.
\end{equation}

We then generate the recommendation by sorting all the nodes and selecting the one, representing either a query or an item, with the highest predicted score.
Formally, we have:
\begin{equation}
\label{eqn:recommend}
a_\mathtt{MAGUS} = \argmax_{a\in\mathcal{A}}\psi_\mathtt{MAGUS}=\argmax_{v\in\mathcal{A}} \widehat{y}_v,
\end{equation}
where $a_\mathtt{MAGUS}$ denotes the query or item to recommend by \textsf{MAGUS}.

For each recommendation $a_\mathtt{MAGUS}$, users are expected to provide real-time feedback.
To efficiently exploit this feedback, we introduce a label propagation-based algorithm to simulate the influence of user feedback on other nodes.
Specially, we update the corresponding node, denoted as $v$, based on the received feedback.
The update function can be written as:
\begin{equation}
\label{eqn:user}
\begin{aligned}
\widehat{y}_v \leftarrow \left\{
\begin{array}{cc}
1, & v\in\mathcal{E}^+,\\
0, & v\in\mathcal{E}^-,\\
\end{array}
\right.
\end{aligned}
\end{equation}
where $v\in\mathcal{E}^+$ denotes the event where $u$ provides positive feedback (e.g., clicks) to $v$, and $v\in\mathcal{E}^-$ denotes the event where $u$ provides negative feedback (e.g., observations without clicks) to $v$.

We then propagate the updated score of $v$ to the remaining nodes, i.e., $v'\in \mathcal{E} \backslash\{v\}$.
In this case, our update function can be written as:
\begin{equation}
\label{eqn:update}
\begin{aligned}
\widehat{y}_{v'} \leftarrow \left\{
\begin{array}{cc}
\min(1, \widehat{y}_{v'} + w_{v'v''}\cdot \widehat{y}_{v''}), & \langle v',v''\rangle \in \overleftarrow{\mathcal{R}}^+,\\
\max(0,\widehat{y}_{v'} - w_{v'v''}\cdot \widehat{y}_{v''}), & \langle v', v''\rangle \in \overleftarrow{\mathcal{R}}^-,\\
\widehat{y}_{v'},& \langle v', v''\rangle \in\mathcal{R}\bot\cup\overrightarrow{\mathcal{R}}^+\cup\overrightarrow{\mathcal{R}}^-.
\end{array}
\right.
\end{aligned}
\end{equation}
We note that, in contrast to Eq.~(\ref{eqn:initial}), which propagates information in the direction of $\overrightarrow{\mathcal{R}}^+\cup\overrightarrow{\mathcal{R}}^-$, the propagation direction in Eq.~(\ref{eqn:update}) is reversed, i.e., $\overleftarrow{\mathcal{R}}_+\cup\overleftarrow{\mathcal{R}}^-$.

After all nodes in $\mathcal{E}$ are updated in each round, we start the next round by successively applying Eqs.~(\ref{eqn:norm}) and (\ref{eqn:recommend}) to generate a new $a_\mathtt{MAGUS}$.  

\subsection{Feature Propagation as Weight Trainer}
\label{sec:weight}
As mentioned in Section~\ref{sec:recommendation}, for each edge in $\mathcal{R}$, we can allocate a learnable weight, when we have access to the offline tuned representation vectors for all the items and users.
Concretely, we denote the representation vector of each item $v\in\mathcal{E}_\mathtt{ITEM}$ as $\bm{e}^{\psi_\mathtt{RE}}_v$, and the representation vector of each user $u\in\mathcal{U}$ as $\bm{e}^{\psi_\mathtt{RE}}_u$.
Since we have access to all the $\bm{e}^{\psi_\mathtt{RE}}_v$ vectors and the sets of browsed items $\mathcal{H}^+_u$s and $\mathcal{H}^-_u$s for all user $u$s, we can proceed with designing a feature propagation algorithm to learn the weights.

Formally, we start by initializing the representation vector of each node $v\in\mathcal{E}$ as follows:
\begin{equation}
\bm{e}_v =
\bm{e}^{\psi_\mathtt{RE}}_v \text{ if } v\in\mathcal{E}_\mathtt{ITEM}, \text{ and }
\bm{e}_v = \bm{e}^\mathtt{INIT}_v \text{ otherwise.}
\end{equation}
where $\bm{e}^\mathtt{INIT}_v$ represents the initialized embedding vector for node $v$, following a given initializer such as \citep{glorot2010understanding}.

We then propagate the feature on the graph.
The update function for each node $v\in\mathcal{E}$ can be formulated as follows:\footnote{We do not explicitly write the propagation layers in Eqs.~(\ref{eqn:feature}) and (\ref{eqn:message}) to make it simple and easy to read. In our experiment, to save the computation cost, we set the number of propagation layers as 1.}
\begin{equation}
\label{eqn:feature}
\bm{e}_v\leftarrow \mathtt{ReLU}\Big(\bm{m}_{v\leftarrow v}+\sum_{v'\in\mathcal{N}^+(v)}\frac{\bm{m}_{v\leftarrow v'}}{|\mathcal{N}^+(v)|}-\sum_{v''\in\mathcal{N}^-(v)}\frac{\bm{m}_{v\leftarrow v''}}{|\mathcal{N}^-(v)|}\Big),
\end{equation}
where $\mathtt{ReLU}(\cdot)$ is an activation function, 
$\mathcal{N}^+(v)$ denotes the set of neighbor nodes connected to $v$ through $\mathcal{R}^+$, and $\mathcal{N}^-(v)$ denotes the set of neighbor nodes connected to $v$ through $\mathcal{R}^-$.
$\bm{m}_{v'\leftarrow v}$ denotes the messages being propagated from node $v$ to node $v'$.
We compute the messages as follows.
\begin{equation}
\label{eqn:message}
\begin{aligned}
\bm{m}_{v\leftarrow v'} = \left\{
\begin{array}{cc}
\bm{W}_1 \cdot \bm{e}_v, & v=v',\\
\bm{W}_1 \cdot \bm{e}_{v'} + \bm{W}_2 \cdot (\bm{e}_v^\top \cdot \bm{e}_{v'})\cdot \bm{e}_v, &  v\neq v',
\end{array}
\right.
\end{aligned}
\end{equation}
where $\bm{W}_\cdot\in\mathbb{R}^{d\times d}$, $d$ is the dimension of the embedding vector of nodes (i.e., $\bm{e}_v$, $v\in\mathcal{E}$).

To supervise the feature propagation, we use $\mathcal{H}^+_u$ and $\mathcal{H}^-_u$ of each user $u$ to establish a log-loss function as follows:
\begin{equation}
\mathcal{L} = -\sum_{u\in\mathcal{U}}\sum_{v\in\mathcal{H}^+_u\cup\mathcal{H}^-_u}\Big(y_v\log\widehat{y}_v+(1-y_v)\log(1-\widehat{y}_v)\Big),
\end{equation}
where $\widehat{y}_v$ is our predicted score relevance for node $v$ regarding user $u$, calculated by $\widehat{y}_v=(\bm{e}^{\psi_\mathtt{RE}}_u)^{\top}\cdot\bm{e}_v$, and $y_v$ is the ground-truth label of node $v$. 
$y_v=1$ if $v\in\mathcal{H}^+_u$, and $y_v=0$ if  $v\in\mathcal{H}^-_u$.

After completing the feature propagation process as described above, we obtain an updated representation vector for each node. 
We then define the weight of each edge in $\mathcal{R}$ based on the similarity between the connected pair of nodes.
This can be expressed as:
\begin{equation}
\label{eqn:trainweight}
w_{vv'}=\bm{e}_v^\top\cdot \bm{e}_{v'} \text{ if }\langle v, v'\rangle\in\mathcal{R}^+\cup\mathcal{R}^-, \text{ and } w_{vv'}=0 \text{ otherwise.} 
\end{equation}

It is worth noting that the supervised learning process described can be conducted offline.
In other words, our relational graph $\mathcal{G}$ can be pre-constructed following \emph{Definition~\ref{def:graph}}, where the edge weights can be pre-computed using either Eq.~(\ref{eqn:weight}) or (\ref{eqn:trainweight}), and these weights be saved within $\mathcal{G}$.
$\mathcal{G}$ can then be stored and only needs to be updated when our recommender system $\psi_\mathtt{RE}(\cdot)$ is updated.
This approach can save computational resources in online operations.

\subsection{System Analysis}
The \textsf{MAGUS} system is summarized in Algorithm~\ref{algo:main} in Appendix~\ref{app:analyze}.
Furthermore, Appendix~\ref{app:analyze} also includes the complexity analysis and discusses the connections to rule-based algorithms, showing that our  \textsf{MAGUS} can be regarded as a hybrid that integrates the structured logic of rule-based methods with the flexibility and adaptability of learning-based techniques.

\section{Experiments}
\label{sec:exp}
% We start our experimental settings and corresponding results with 4 questions and use them to lead the following discussions.
% \begin{itemize}[topsep = 3pt,leftmargin =5pt]
% \item \textbf{[Q1]} Does the inclusion of ``query'' truly enhance item recommendations?
% \item \textbf{[Q2]} Does our graph structure effectively organize the relationship among queries and items?
% \item \textbf{[Q3]} What is the impact of using different multiple-round recommendation settings for \textsf{MAGUS}?
% \item \textbf{[Q4]} Is \textsf{MAGUS} easy-to-deploy in practice? 
% \end{itemize}

\begin{table*}[t]
	\centering
	\caption{Results comparison of items recommendations in terms of SAC, and joint recommendations of both queries and items in terms of RA@3, SA@3, and SA@5.
    Since SAC metric measures the performance on the single-round item recommendation task, we do not report SAC for \textsf{MAGUS} and \textsf{MAGUS}$^+$.
	* indicates $p < 0.001$ in significance tests compared to the best baseline.
	}
	\vspace{-3mm}
	\resizebox{0.95\textwidth}{!}{
		\begin{tabular}{@{\extracolsep{4pt}}|c|cccc|cccc|cccc|}
		\toprule
			\multirow{2}{*}{Methods} & \multicolumn{4}{c}{Amazon} |& \multicolumn{4}{c}{Alipay} |& \multicolumn{4}{c}{Tmall}|\\
			\cmidrule{2-5}
			\cmidrule{6-9}
			\cmidrule{10-13}
			{} & SAC & RA@3 & SA@3 & SA@5 & SAC & RA@3 & SA@3 & SA@5 & SAC & RA@3 & SA@3 & SA@5\\
			\midrule
			MPS & 
			0.332 & 0.612 & 0.174 & 0.255 & 
			0.298 & 0.541 & 0.125 & 0.181 & 
			0.312 & 0.575 & 0.164 & 0.208 \\
                \midrule
			Hybrid & 
			0.394 & 0.665 & 0.312 & 0.406 & 
			0.365 & 0.592 & 0.286 & 0.345 & 
			0.344 & 0.592 & 0.295 & 0.337 \\
			\midrule
			FM & 
			0.634 & 0.773 & 0.672 & 0.757 & 
			0.716 & 0.815 & 0.767 & 0.846 & 
			0.718 & 0.832 & 0.745 & 0.824 \\
                \midrule
                FM+CRM & 
                / & 0.787 & 0.675 & 0.760 & 
			/ & 0.826 & 0.798 & 0.867 & 
			/ & 0.880 & 0.771 & 0.852 \\
                \midrule
                FM+ME & 
                / & 0.794 & 0.688 & 0.771 & 
			/ & 0.817 & 0.789 & 0.860 & 
			/ & 0.847 & 0.754 & 0.831 \\
                \midrule
                FM+EAR & 
                / & 0.795 & 0.695 & 0.769 & 
			/ & 0.825 & 0.796 & 0.866 & 
			/ & 0.878 & 0.765 & 0.850 \\
                \midrule
			FM+\textsf{MAGUS} &
			/ & \textbf{0.816}$^{*}$ & \textbf{0.742}$^{*}$ & \textbf{0.798}$^{*}$ & 
			/ & \textbf{0.843}$^{*}$ & \textbf{0.825}$^{*}$ & \textbf{0.888}$^{*}$ & 
			/ & \textbf{0.894}$^{*}$ & \textbf{0.791}$^{*}$ & \textbf{0.877}$^{*}$ \\
                \midrule
			DeepFM & 
			0.676 & 0.784 & 0.693 & 0.798 & 
			0.730 & 0.825 & 0.787 & 0.875 & 
			0.729 & 0.843 & 0.766 & 0.841 \\
                \midrule
                DeepFM+CRM & 
                / & 0.796 & 0.705 & 0.805 & 
			/ & 0.840 & 0.817 & 0.882 & 
			/ & 0.879 & 0.802 & 0.881 \\
                \midrule
                DeepFM+ME & 
                / & 0.795 & 0.698 & 0.794 & 
			/ & 0.835 & 0.811 & 0.879 & 
			/ & 0.856 & 0.775 & 0.864 \\
                \midrule
                DeepFM+EAR & 
                / & 0.810 & 0.743 & 0.807 & 
			/ & 0.839 & 0.818 & 0.884 & 
			/ & 0.885 & 0.800 & 0.885 \\
                \midrule
			DeepFM+\textsf{MAGUS} &
			/ & \textbf{0.833}$^{*}$ & \textbf{0.767}$^{*}$ & \textbf{0.811}$^{*}$ & 
			/ & \textbf{0.851}$^{*}$ & \textbf{0.832}$^{*}$ & \textbf{0.895}$^{*}$ & 
			/ & \textbf{0.903}$^{*}$ & \textbf{0.814}$^{*}$ & \textbf{0.892}$^{*}$ \\
			\midrule
			PNN & 
			0.688 & 0.788 & 0.690 & 0.792 & 
			0.741 & 0.833 & 0.775 & 0.870 & 
			0.722 & 0.823 & 0.753 & 0.831 \\
                \midrule
                PNN+CRM & 
                / & 0.807 & 0.714 & 0.798 & 
			/ & 0.851 & 0.844 & 0.899 & 
			/ & 0.870 & 0.798 & 0.827 \\
                \midrule
                PNN+ME & 
                / & 0.813 & 0.749 & 0.805 & 
			/ & 0.845 & 0.820 & 0.884 & 
			/ & 0.855 & 0.776 & 0.845 \\
                \midrule
                PNN+EAR & 
                / & 0.814 & 0.747 & 0.802 & 
			/ & 0.853 & 0.845 & 0.898 & 
			/ & 0.872 & 0.801 & 0.863 \\
                \midrule
			PNN+\textsf{MAGUS} & 
			/ & \textbf{0.839}$^{*}$ & \textbf{0.772}$^{*}$ & \textbf{0.817}$^{*}$ & 
			/ & \textbf{0.865}$^{*}$ & \textbf{0.852}$^{*}$ & \textbf{0.911}$^{*}$ & 
			/ & \textbf{0.884}$^{*}$ & \textbf{0.812}$^{*}$ & \textbf{0.876}$^{*}$ \\
			\midrule
			MMoE &
			0.631 & 0.770 & 0.663 & 0.744 & 
			0.703 & 0.802 & 0.745 & 0.811 & 
			0.723 & 0.842 & 0.752 & 0.830 \\
                \midrule
			MMoE+\textsf{MAGUS} &
			/ & \textbf{0.801}$^{*}$ & \textbf{0.725}$^{*}$ & \textbf{0.776}$^{*}$ & 
			/ & \textbf{0.833}$^{*}$ & \textbf{0.820}$^{*}$ & \textbf{0.876}$^{*}$ & 
			/ & \textbf{0.898}$^{*}$ & \textbf{0.802}$^{*}$ & \textbf{0.881}$^{*}$  \\
			\midrule
			DIN &
			0.697 & 0.798 & 0.696 & 0.813 & 
			0.757 & 0.845 & 0.793 & 0.886 & 
			0.736 & 0.855 & 0.774 & 0.848 \\
                \midrule
			DIN+\textsf{MAGUS} &
			/ & \textbf{0.845}$^{*}$ & \textbf{0.775}$^{*}$ & \textbf{0.828}$^{*}$ & 
			/ & \textbf{0.878}$^{*}$ & \textbf{0.865}$^{*}$ & \textbf{0.922}$^{*}$ & 
			/ & \textbf{0.904}$^{*}$ & \textbf{0.818}$^{*}$ & \textbf{0.902}$^{*}$ \\
			\midrule
			LSTM &
			0.692 & 0.789 & 0.690 & 0.808 & 
			0.752 & 0.840 & 0.782 & 0.876 & 
			0.728 & 0.846 & 0.759 & 0.837 \\
                \midrule
			LSTM+\textsf{MAGUS} &
			/ & \textbf{0.840}$^{*}$ & \textbf{0.773}$^{*}$ & \textbf{0.821}$^{*}$ & 
			/ & \textbf{0.870}$^{*}$ & \textbf{0.861}$^{*}$ & \textbf{0.918}$^{*}$ & 
			/ & \textbf{0.901}$^{*}$ & \textbf{0.808}$^{*}$ & \textbf{0.892}$^{*}$ \\
			\midrule
			GRU &
			0.707 & 0.803 & 0.699 & 0.818 & 
			0.762 & 0.848 & 0.799 & 0.889 & 
			0.732 & 0.852 & 0.771 & 0.845 \\
                \midrule
			GRU+\textsf{MAGUS} &
			/ & \textbf{0.848}$^{*}$ & \textbf{0.788}$^{*}$ & \textbf{0.831}$^{*}$ & 
			/ & \textbf{0.882}$^{*}$ & \textbf{0.871}$^{*}$ & \textbf{0.926}$^{*}$ & 
			/ & \textbf{0.909}$^{*}$ & \textbf{0.821}$^{*}$ & \textbf{0.901}$^{*}$ \\
                \midrule
			RGCN &
			0.668 & 0.781 & 0.687 & 0.784 & 
			0.736 & 0.828 & 0.785 & 0.877 & 
			0.722 & 0.828 & 0.747 & 0.825 \\
                \midrule
			RGCN+\textsf{MAGUS} &
			/ & \textbf{0.841}$^{*}$ & \textbf{0.775}$^{*}$ & \textbf{0.824}$^{*}$ & 
			/ & \textbf{0.873}$^{*}$ & \textbf{0.860}$^{*}$ & \textbf{0.912}$^{*}$ & 
			/ & \textbf{0.897}$^{*}$ & \textbf{0.810}$^{*}$ & \textbf{0.893}$^{*}$ \\
                \midrule
			RGCN+\textsf{MAGUS}$^+$ & 
			/ & \textbf{0.852}$^{*}$ & \textbf{0.787}$^{*}$ & \textbf{0.831}$^{*}$ & 
			/ & \textbf{0.882}$^{*}$ & \textbf{0.870}$^{*}$ & \textbf{0.925}$^{*}$ & 
			/ & \textbf{0.903}$^{*}$ & \textbf{0.820}$^{*}$ & \textbf{0.902}$^{*}$ \\ 
			\midrule
			RGAT &
			0.675 & 0.785 & 0.695 & 0.794 & 
			0.748 & 0.838 & 0.782 & 0.878 & 
			0.730 & 0.843 & 0.762 & 0.839 \\
                \midrule
			RGAT+\textsf{MAGUS} &
			/ & \textbf{0.850}$^{*}$ & \textbf{0.785}$^{*}$ & \textbf{0.828}$^{*}$ & 
			/ & \textbf{0.878}$^{*}$ & \textbf{0.868}$^{*}$ & \textbf{0.921}$^{*}$ & 
			/ & \textbf{0.905}$^{*}$ & \textbf{0.817}$^{*}$ & \textbf{0.897}$^{*}$ \\
			\midrule
			RGAT+\textsf{MAGUS}$^+$ & 
			/ & \textbf{0.867}$^{*}$ & \textbf{0.798}$^{*}$ & \textbf{0.840}$^{*}$ & 
			/ & \textbf{0.891}$^{*}$ & \textbf{0.879}$^{*}$ & \textbf{0.933}$^{*}$ & 
			/ & \textbf{0.911}$^{*}$ & \textbf{0.826}$^{*}$ & \textbf{0.914}$^{*}$ \\ 
                \midrule
			GIPA &
			0.688 & 0.798 & 0.707 & 0.799 & 
			0.756 & 0.847 & 0.796 & 0.885 & 
			0.751 & 0.849 & 0.778 & 0.843 \\
                \midrule
			GIPA+\textsf{MAGUS} &
			/ & \textbf{0.856}$^{*}$ & \textbf{0.798}$^{*}$ & \textbf{0.834}$^{*}$ & 
			/ & \textbf{0.877}$^{*}$ & \textbf{0.867}$^{*}$ & \textbf{0.918}$^{*}$ & 
			/ & \textbf{0.912}$^{*}$ & \textbf{0.824}$^{*}$ & \textbf{0.902}$^{*}$ \\
			\midrule
			GIPA+\textsf{MAGUS}$^+$ & 
			/ & \textbf{0.881}$^{*}$ & \textbf{0.785}$^{*}$ & \textbf{0.849}$^{*}$ & 
			/ & \textbf{0.892}$^{*}$ & \textbf{0.881}$^{*}$ & \textbf{0.934}$^{*}$ & 
			/ & \textbf{0.919}$^{*}$ & \textbf{0.832}$^{*}$ & \textbf{0.918}$^{*}$ \\
			\bottomrule
		\end{tabular}
	}
	\label{tab:res}
	\vspace{-2mm}
\end{table*}

\subsection{Experimental Settings}
\label{sec:exp}
\minisection{Dataset Description.}
We conduct extensive experiments on 3 industrial real-world e-commerce datasets, namely Amazon\footnote{\url{https://jmcauley.ucsd.edu/data/amazon/}}, Alipay\footnote{\url{https://tianchi.aliyun.com/dataset/dataDetail?dataId=53}}, and Tmall\footnote{\url{https://tianchi.aliyun.com/dataset/dataDetail?dataId=42}}.
For each dataset, we collect user-item interaction histories (i.e., $\mathcal{H}^+_u \cup \mathcal{H}^-_u$) of users $u\in\mathcal{U}$, and gather all items to form $\mathcal{V}$.
In the process of constructing the query set $\mathcal{Q}$, the pivotal step is the assembly of the word set $\mathcal{W}$, and one can then derive $\mathcal{Q}$ following \emph{Definition~\ref{def:graph}}.
Concretely, we initiate the word set $\mathcal{W}$ by extracting words from the categorical features of items in $\mathcal{V}$.
Let's take the Amazon dataset as an example.
If an item has the brand \texttt{Coxlures} and belongs to the categories \texttt{Sports} and \texttt{Dance}, we generate words like \texttt{coxlures}, \texttt{sports}, and \texttt{dance}.
By iterating through all items in $\mathcal{V}$m we compile the set $\mathcal{W}$.
For each dataset, we perform a temporal split, dividing it into training, validation, and test sets in a ratio of 6:2:2 based on time steps.
We exclude sequences with a length of less than 30 or those that lack items with positive feedback.
We then randomly select 30 items to form a session. 
During the random selection of items, we ensure that at least one item with positive feedback is included.
We provide comprehensive details about the datasets, the data prepossessing, and the relational graph construction in Appendix~\ref{app:dataset}, \ref{app:preprocess}, \ref{app:graph}.

\minisection{Baseline Description.}
Although our \textsf{MAGUS} framework can enable any recommendation method to recommend queries or items in a multiple-round setting, the ultimate goal, as formulated in \emph{Definition~\ref{def:task}}, is to identify an item satisfying user needs in each user session.
In this context, we introduce 12 recommendation methods as base recommender systems, and our \textsf{MAGUS} is evaluated in an ablation style.
These recommendation methods include 
MPS (Most Popular Suggestion) \citep{dehghani2017learning,sordoni2015hierarchical},
Hybrid (Hybrid Suggestion) \citep{bar2011context},
FM (Factorization Machine) \citep{rendle2010factorization},
DeepFM \citep{guo2017deepfm},
PNN (Product-based Neural Network) \citep{qu2016product},
MMoE (Multi-gate Mixture-of-Experts) \citep{ma2018modeling},
DIN (Deep Interest Network) \citep{zhou2018deep},
LSTM (Long-Short Term Memory) \citep{hochreiter1997long},
GRU (Gated Recurrent Unit) \citep{hidasi2015session},
RGCN (Relational Graph Convolutional Network) \citep{schlichtkrull2018modeling},
RGAT (Relational Graph Attention Network) \citep{busbridge2019relational}, and
GIPA (General Information Propagation Algorithm) \citep{zheng2021gipa}.
Regarding the proposed multiple-round recommendation task for both queries and items, there are currently no existing recommendation solutions tailored specifically for this task. 
% Therefore, we adopt the following two types of adjustments.
% One type of adjustment pertains to the incorporation of user responses in the masonry layout. 
we extend the action space from considering only items $\mathcal{V}$ to encompass both items and queries $\mathcal{V}\cup\mathcal{Q}$.
We introduce a recurrent neural network to exploit user responses during multiple-round interactions. 
In this approach, the final embedding layer of the single-round baselines is fed into a recurrent unit, allowing item information and corresponding user feedback from each round to be carried forward to subsequent rounds.
A detailed description of the above adjustment can be found in Appendix~\ref{app:baseline}.
% To apply the above baselines to our setting, the following two adjustments are made.
% One is to extend the action space from considering only items $\mathcal{V}$ to encompass both items and queries  $\mathcal{V}\cup\mathcal{Q}$.
% The other one is incorporating user response.
% To enable these methods to utilize user responses during multiple-round interactions, a recurrent neural network is introduced, where the last embedding layer of these single-round baselines is fed into a unit, allowing information from each round to be carried forward to subsequent rounds.

Another way to enable the base recommendation methods to fit the multiple-round recommendation task is through the development of conversational recommender systems \citep{sun2018conversational,lei2020estimation}.
% , which enable the recommendation method to establish multiple-round interactions with users in the conversation layout.
To fit these approaches into our joint recommendation task for items and queries, we let the set of nodes in our relational graph represent the action space of the conversational recommendation methods, including ME (Max-Entropy strategy that always queries the nodes with the maximum entropy in terms of all candidate items), CRM (Conversational Recommender Model) \citep{sun2018conversational}, and EAR (Estimation-Action-Reflection) \citep{lei2020estimation}.

As detailed in Algorithm~\ref{algo:main} (in Appendix~\ref{app:analyze}), there are two versions of the \textsf{MAGUS} system, each employing a different method to calculate the edge weights of the relational graph (i.e., line~\ref{line:weight}).
% We can summarize these versions as follows.
\textsf{MAGUS} is the \textsf{MAGUS} system computing edge weights using Eq.~(\ref{eqn:weight}). 
\textsf{MAGUS}$^+$ is a variant of the \textsf{MAGUS} system where the edge weights are calculated using Eq.~(\ref{eqn:trainweight}).

We use notations such as X to represent instances where we apply the first type of adjustment to method X, and X + ME, X + CRM, X + EAR for the second type of adjustment.
Additionally, X+\textsf{MAGUS} and X+\textsf{MAGUS}$^+$ represent instances where we employ method X as $\psi_\mathtt{RE}(\cdot)$.
% within Algorithm~\ref{algo:main}.
Here, X can represent one of the following recommendation bases: FM, DeepFM, PNN, MMoE, DIN, LSTM, GRU, RGCN, RGAT, and GIPA.

\minisection{Simulation Description.}
To accomplish the loop between humans and the system, we introduce a novel multiple-round recommendation simulator supporting both query and item recommendations.
This simulator comprises two key components: the user agent and the recommender agent.
In this setup, the recommender agent has the option to either select an item or form a query during each round of interaction.
The response generated by the user agent hinges on whether the recommendation $a_\mathtt{MAGUS}$ corresponds to a target item or a query that is part of a target item. 
If it does, the user agent provides a positive response; otherwise, it issues a negative response.
We choose to set the length of the recommendation lists as 3 (denoted as $N$).
We also establish a predefined limit on the number of interaction rounds (denoted as $K_\mathtt{MAX}$), which is explicitly defined within the evaluation metrics to be introduced subsequently.

% to use the recommendation lists of length 3 (denoted as $N$); and the maximum number of rounds for evaluation  is specified in the evaluation metrics introduced later.
% A session concludes when the recommender agent successfully recommends a target item to the user agent or the number of rounds reaches $K_\mathtt{MAX}$.

In order to expand the applicability and versatility of the \textsf{MAGUS} system, we have implemented an option to seamlessly integrate a large language model with a plug-and-play architecture.
This innovative enhancement equips \textsf{MAGUS} with the capability to engage in natural language communication with human users.
A detailed description of our simulator and the integration of the large language models are provided in Appendix~\ref{app:agent}, \ref{app:llm}.

% To broaden the application scope of \textsf{MAGUS}, we also offer an option to integrate a plug-and-play large language model.
% This enhancement enables our \textsf{MAGUS} to communicate with humans in natural language, thus improving user-friendliness.

\minisection{Evaluation Metrics.}
In our evaluation, we focus on two primary types of metrics in effectively identifying items that match user needs within a session.
One is round-wise accuracy, denoted as RA@$K_\mathtt{MAX}$, which evaluates the recommendation performance during each round of interaction.
For every round $k$, if a user clicks on the recommendation at position $b_k$, we calculate  RA@$K_\mathtt{MAX}$ using RA@$K_\mathtt{MAX}=1/\log_2(b_k+1)$.
If a user does not click on any recommendation during a round, RA@$K_\mathtt{MAX}$ is set to 0.
The final RA@$K_\mathtt{MAX}$ is determined as the average of all the RA@$K_\mathtt{MAX}$ values.
The other one is session-wise accuracy, denoted as SA@$K_\mathtt{MAX}$, which assesses whether the system successfully recommends a target item within $K_\mathtt{MAX}$ interactions in each session.
If the system succeeds in recommending a target item, SA@$K_\mathtt{MAX}$ is set to 1 for that session; otherwise, it is set to 0.
The final SA@$K_\mathtt{MAX}$ is computed as the average of all the SA@$K_\mathtt{MAX}$ values.

Additionally, to gauge the impact of combining query and item recommendations, we provide a single-round accuracy (denoted as SAC) metric for each baseline method.
In each session, SAC is set to 1 if the recommendation list contains at least one target item, and 0 otherwise.
The final SAC is determined as the average of all the SAC values.

Implementation details along with the code link are available in Appendix~\ref{app:code}.
% We also provide the implementation details in the supplementary materials.

\begin{figure}[t]
	\centering
	\vspace{-2mm}
	\includegraphics[width=1.00\linewidth]{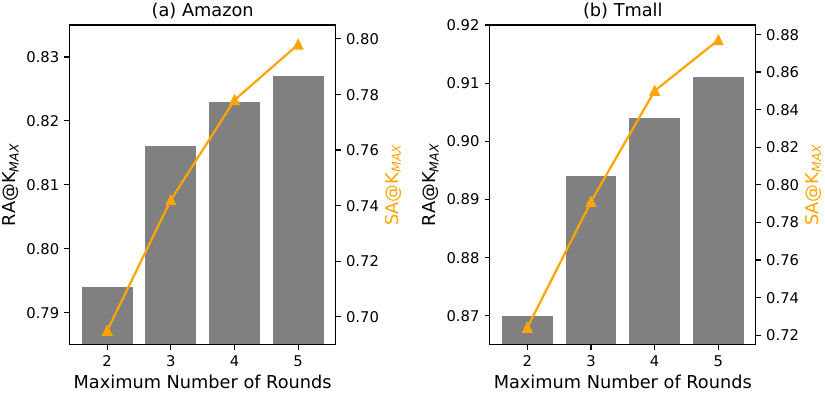}
	\vspace{-7mm}
	\caption{Performance comparisons of \textsf{MAGUS} with different maximum numbers of rounds (i.e., $K_\mathtt{MAX}$) in terms of RA@$K_\mathtt{MAX}$ and SA@$K_\mathtt{MAX}$ with the FM recommendation base.}
	\label{fig:round}
	\vspace{-4mm}
\end{figure}

% \minisection{Empowering our Simulator with Large Language Models.}
% A detailed description of this integration is provided in Appendix~\ref{app:llm}.
% Integrating the \textsf{MAGUS} system into an existing recommendation platform should not introduce a significant computational burden, since the primary modifications only involve organizing the relational graph and updating the graph by user responses.
% Here, a potential issue is that queries formed by the \textsf{MAGUS} system may occasionally manifest as somewhat artificial, potentially diverging from user-friendly formulations.
% In response to this challenge, a viable approach is to harness the capabilities of large language models while employing certain prompts to steer the generation of queries toward user-friendliness.

% We provide the simulator and the associated code at \url{https://anonymous.4open.science/r/MAGUS-CCF4}.

\subsection{Performance Comparisons}
We present the outcomes of our experiments in Table~\ref{tab:res}.
We summarize our findings as follows.

\minisection{Performance Comparisons between Jointly using Queries and Items and Solely using Items.}
As our SAC is evaluating the recommendation performance of solely considering items, we compare the results of single-round performance SAC against multiple-round performance RA@3, SA@3, and SA@5 of the base recommendation method (i.e., rows of FM, DeepFM, PNN, MMoE, DIN, LSTM, GRU, RGCN, RGAT, and GIPA).
These results show that combining items and queries in item recommendations can significantly improve the recommendation performance across 12 diverse recommendation approaches on all 3 datasets.
This improvement also would be attributed to our multiple-round setting, which allows real-time user feedback and better aligns with user preferences.
Notably, popularity-based methods like MPS and Hybrid perform well in terms of RA@3 but struggle with SA@3 and SA@5, likely because high popularity often occurs in queries, leading to an overemphasis on recommending queries.

\minisection{Performance Comparisons between \textsf{MAGUS} (denoted as \textsf{MAGUS} + X) and using Sequential Neural Networks as Update Formula (denoted as X).}
As described in Section~\ref{sec:exp}, we extend the existing single-round recommendation methods (i.e., the base recommendation methods) to the multiple-round setting by a recurrent neural network based framework.
When comparing X+\textsf{MAGUS} and X in terms of RA@3, SA@3, and SA@5, one can see that X+\textsf{MAGUS} significantly outperforms X.
This demonstrates that \textsf{MAGUS} provides an effective solution for jointly considering queries and items. 
One possible explanation for the improved performance of \textsf{MAGUS} is its capability to leverage the dependence among queries and items through the label propagation algorithm.
This enables the recommender system to encode user feedback over time, resulting in more accurate recommendations.

\begin{figure}[t]
	\centering
	\vspace{-2mm}
	\includegraphics[width=1.00\linewidth]{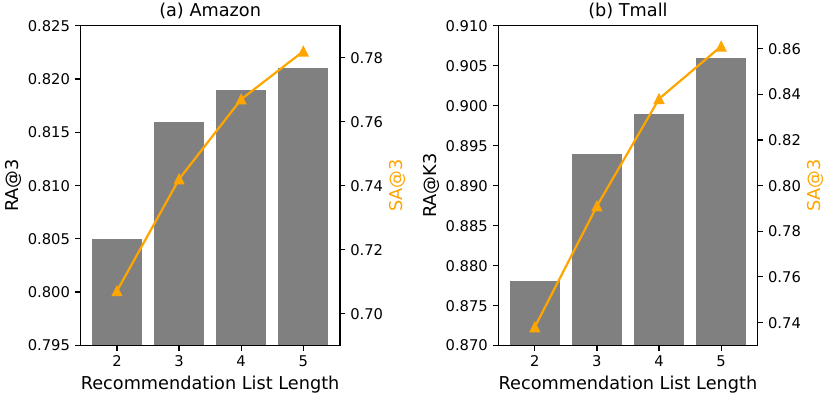}
	\vspace{-7mm}
	\caption{Performance comparisons of \textsf{MAGUS} using different lengths of the recommendation list in terms of RA@3 and SA@3 with the FM recommendation base.}
	\label{fig:len}
	\vspace{-3mm}
\end{figure}

\minisection{Performance Comparisons between \textsf{MAGUS} (denoted as X + \textsf{MAGUS}) and using Conversational Recommender Systems (denoted as X + CRM, X + ME, X + EAR).}
As described in Section~\ref{sec:exp}, we incorporate the existing conversational recommender systems into the base recommendation methods FM, DeepFM, and PNN.
When comparing X+\textsf{MAGUS} and X + CRM, X + ME, X + EAR in terms of RA@3, SA@3, and SA@5, one can see that X+\textsf{MAGUS} can consistently achieve better performance.
This result can be attributed to several key differences between \textsf{MAGUS} and the conversational methods, summarized as follows:
(i) Unlike the \textsf{MAGUS} system, conversational recommendation methods (along with their simulations in the experimental setting) do not take query information into account.
(ii) Compared to the \textsf{MAGUS} method, conversational recommendation methods do not explicitly model the rich interdependence among attributes (where attributes correspond to words and word combinations in the \textsf{MAGUS} method).
(iii) Many online recommendation methods utilize reinforcement learning techniques. 
While reinforcement learning methods can handle complex states and encode multiple factors, they often face challenges such as the need for a large number of training samples (referred to as the data insufficiency issue) and relatively small action space.

\minisection{Performance Comparisons between Graph-based Recommendations and other Recommendations.}
When considering SAC, sequential recommendation methods such as GRU, LSTM, and DIN often outperform other baseline methods.
However, concerning RA@3, SA@3, and SA@5, the graph-based methods, namely RGCN, RGAT, and GIPA, achieve comparable or superior results compared to the sequential recommendation methods.
This suggests that graph-based methods, when applied to our relational graph, can effectively exploit the dependence among queries and items to make recommendations.
This finding also supports the advantage of using a relational graph to bridge queries and items.

\minisection{Performance Comparisons between \textsf{MAGUS} and \textsf{MAGUS}$^+$.}
When comparing X+\textsf{MAGUS}$^+$ and X+\textsf{MAGUS} (where X is RGCN, RGAT, and GIPA), we can observe that \textsf{MAGUS}$^+$ can bring consistent enhancements, which indicates that integrating these learned edge weights into the multiple-round recommendation process can have a positive influence.
This can explained as the learned edge weights would be capable of capturing more meaningful relationships between queries and items, which also verifies the superiority of using our relational graph.

\subsection{Hyperparameter Study}
\label{sec:ablation}

\minisection{Impact of Maximum Number of Rounds $K_\mathtt{MAX}$.}
One important hyper-parameter in \textsf{MAGUS} is the maximum number of rounds, denoted as $K_\mathtt{MAX}$.
We examined how the performance of \textsf{MAGUS} changes when setting $K_\mathtt{MAX}$ to different values, i.e., $K_\mathtt{MAX}=2,3,4,5$ on the Amazon and the Tmall datasets.
Results depicted in Figure~\ref{fig:round} consistently indicate that multiple-round recommendations with more feedback rounds from users outperform those with fewer rounds.
This highlights the advantages of implementing a multiple-round recommendation service.

\minisection{Impact of Recommendation Length $N$.}
The length of the recommendation list at each round (denoted as $N$) is another crucial hyper-parameter in \textsf{MAGUS}.
To investigate its impact, we experimented with different values, i.e., $N=2,3,4,5$, on the Amazon and the Tmall datasets.
Results replayed in Figure~\ref{fig:len} show that increasing the length of the recommendation list can lead to improved performance.
This improvement can be attributed to the fact that a larger list provides users with more choices and options to choose from.
However, it is worth noting that the performance gain achieved by increasing the length of recommendation list $N$ is not as substantial as that obtained by increasing the number of rounds $K_\mathtt{MAX}$.
This is because, compared to $N$ recommendations in one round, recursively recommending one item in $N$ rounds could enable the model to update the predicted scores to benefit the following recommendations, leading to a significant impact on overall performance.

\subsection{Study for Deployment Feasibility}
We present a robustness study that incorporates the analysis of ambiguous user feedback within our system in Appendix~\ref{app:robustness}.
We also derive into a complexity study in Appendix~\ref{app:time}.
Furthermore, we offer detailed case studies in Appendix~\ref{app:case}.
We then discuss the deployment architecture, the extension from top-1 to top-N recommendation, and the use case for exploratory research in Appendix~\ref{app:arch}, \ref{app:topn}, \ref{app:explore}.

\section{Conclusion and Future Work}
In this paper, we propose a novel recommendation system \textsf{MAGUS} of jointly considering queries and items.
% where we construct a relational graph and devise a label propagation algorithm designed to encapsulate the interdependence between queries and items. 
Importantly, the \textsf{MAGUS} framework can be seamlessly integrated into existing recommendation platforms. 
In the future, it would be interesting to evaluate and deploy the  \textsf{MAGUS} system on an online e-commerce platform.

% \clearpage
\bibliographystyle{ACM-Reference-Format}
% \balance
\bibliography{main}
\clearpage
\appendix
\section{A Summary of Notations}
\label{app:notation}
\begin{table}[h]
    \vspace{-2mm}
	\caption{A summary of notations. 
		% Uppercase letters (e.g.,~$C$,$R$,$O$,$X$,$Y$,$\hat{Y}$) denote random variables, and lowercase letters (e.g.,~$c,r,o,\bm{x},y,\hat{y}$) denote the corresponding value for each data point.
	}  
	\label{tab:notation}
	\vspace{-4mm}
	\begin{center}  
		\begin{tabular}{p{2.5cm}<{\centering}|p{5cm}<{\centering}}
			\toprule
			\textbf{Notations} & \textbf{Explanations} \\
			\midrule
			$u\in\mathcal{U}, v\in\mathcal{V}$ & user $u$ in set $\mathcal{U}$, item $v$ in set $\mathcal{V}$ \\
			\midrule
			$q\in\mathcal{Q},w\in\mathcal{W}$ & query $q$ in set $\mathcal{Q}$, word $w$ in set $\mathcal{W}$ \\
			\midrule
			$a\in\mathcal{A}$ & query or item $a$ in set $\mathcal{A}=\mathcal{V}\cup\mathcal{Q}$ \\
                \midrule
			$\mathcal{G}=(\mathcal{E},\mathcal{R})$ & node set $\mathcal{E}$ in graph $\mathcal{G}$ contains both individual words $\mathcal{W}$ and certain combinations of words representing queries $\mathcal{Q}$ and items $\mathcal{V}$\\
			\midrule		$\mathcal{R}=\mathcal{R}^{+}\cup\mathcal{R}^{-} \cup\mathcal{R}^{\bot}$ & edge set $\mathcal{R}$ in graph $\mathcal{G}$ contains three set of relationships between nodes\\
			\midrule
			$\mathcal{H}_u^+, \mathcal{H}_u^-$ & set of items receiving positive, negative feedback from user $u$\\
                \midrule
			$\mathcal{H}_u^q$ & set of searched queries of user $u$\\
			\midrule
			$\bm{e}_v, \widehat{y}_v, \psi_\mathtt{RE}(v)$ & embedding vector, updated score, predicted score of node $v\in\mathcal{E}$ \\
			\bottomrule
		\end{tabular}  
	\end{center} 
	\vspace{-3mm}
\end{table}

\section{System Analysis}
\label{app:analyze}
\subsection{Overall Algorithm}
The \textsf{MAGUS} system is summarized in Algorithm~\ref{algo:main}.
Lines~\ref{line:recommend}, \ref{line:graph}, \ref{line:weight} and \ref{line:offlineupdate} represent the offline training part, while lines 4 to 12 are the online update part.
As mentioned earlier, the offline training part could be pre-computed and stored, making the online interaction process more efficient.

\begin{algorithm}[h]
	\caption{The \textsf{MAGUS} System}
	\label{algo:main}
	\begin{algorithmic}[1]
		\REQUIRE
            positive and negative browsed items for all users $\{\mathcal{H}^+_u|u\in\mathcal{U}\}$ and  $\{\mathcal{H}^-_u|u\in\mathcal{U}\}$;
            optional: searched queries for all users $\{\mathcal{H}^q_u|u\in\mathcal{U}\}$. 
		% any offline tuned recommender system $\psi_\mathtt{RE}(\cdot)$, maximum number of rounds $K_\mathtt{MAX}$; optional: embedding vectors of all the items $\{\bm{e}^\mathtt{RE}_v|v\in\mathcal{E}_\mathtt{ITEM}\}$ and all the users $\{\bm{e}^\mathtt{RE}_u|u\in\mathcal{U}\}$,  
		\ENSURE
		recommended query or item $a_\mathtt{MAGUS}$ at each round.
		\vspace{1mm}
            \STATE Offline train a recommendation model $\psi_\mathtt{RE}(\cdot)$ upon $\mathcal{H}^+_u$s and $\mathcal{H}^-_u$s for $u\in\mathcal{U}$.
            \label{line:recommend}
            \STATE Offline build a relational graph $\mathcal{G}=(\mathcal{E},\mathcal{R})$ by \emph{Definition~\ref{def:graph}}.
            \label{line:graph}
            \STATE Offline compute the weights of the edges in $\mathcal{R}$ using Eq.~(\ref{eqn:weight}) or Eq.~(\ref{eqn:trainweight}). 
            \label{line:weight}
            \FOR{each online session for user $u$}
            \STATE Initialize $k=0$.
            \label{line:onlinebegin}
            \STATE Initialize the scores of all nodes using Eqs.~(\ref{eqn:recommendation}), (\ref{eqn:initial}), and (\ref{eqn:query}).
            \REPEAT
            \STATE Normalize the scores of all nodes using Eq.~(\ref{eqn:norm}).
            \STATE Compute $a_\mathtt{MAGUS}$ using Eq.~
            (\ref{eqn:recommend}).
            \label{line:1}
            \STATE Recommend $a_\mathtt{MAGUS}$ and receive corresponding response.
            \label{line:2}
            \STATE Update the scores of the nodes using Eqs.~(\ref{eqn:user}) and (\ref{eqn:update}).
             \label{line:3}
            \STATE Go to next round: $k\leftarrow k+1$.
            \UNTIL{$a_\mathtt{MAGUS}\in\mathcal{V}_\mathtt{TARGET}$ or $k>K_\mathtt{MAX}$.}
            \STATE Collect session data into $\mathcal{H}^+_u$ and $\mathcal{H}^-_u$.
            \label{line:onlineend}
            \ENDFOR
		\STATE Update $\psi_\mathtt{RE}(\cdot)$ using data in new $\mathcal{H}^+_u$s and new $\mathcal{H}^-_u$s.
        \label{line:offlineupdate}
	\end{algorithmic}
\end{algorithm}

Our approach can be regarded as a combination of non-parametric recommendation methods relying on connections between queries and items, and parametric recommendation methods based on user browsing logs.   
Concretely, our recommendation problem can be formulated as a node classification problem on our graph $\mathcal{G}$, aiming to recommend a node to a given user. 
To update our predictions in each $k$-th round of interaction, we use the formula $\bm{Y}^{(k+1)}=\Theta \bm{S}\bm{Y}^{(k)}+\bm{Y}^{(0)}$, where $\bm{S}$ is the symmetric normalized version of the adjacency matrix $\bm{A}$ given by $\bm{S}=\bm{D}^{-\frac{1}{2}}\bm{A}\bm{D}^{-\frac{1}{2}}$ and $\bm{D}$ represents the degree matrix of the graph $\mathcal{G}$, and $\Theta$ is a learnable weight (as described in Section~\ref{sec:weight}).
Here, our adjacency matrix $\bm{A}$ incorporates the connections among queries and items, as introduced in \emph{Definition~\ref{def:graph}}; whereas $\bm{Y}^{(0)}$ and $\Theta$ are computed using an offline learned recommendation method.
As a result, our \textsf{MAGUS} system exhibits superior generalization and greater capacity for managing intricate connections between queries and items, compared to either purely parametric or non-parametric recommendation methods.

% In contrast, rule-based methods have better interpretation ability, making it easier to infer why the system acts in a certain way than with the MAGUS method.

\subsection{Complexity Analysis}
The \textsf{MAGUS} system consists of two key components: the feature propagation part during offline training and the label propagation part during online inference.
Matrix multiplication on graph, which is a fundamental operation in these components, has a computational complexity of $O(|\bm{S}_+|\cdot d^2)$ as stated in \citep{wang2019neural}, where $|\bm{S}_+|$ is the number of nonzero entities in $\bm{S}$.
% $\bm{L}=\bm{D}^{-1/2}\bm{A}\bm{D}^{-1/2}$ where $\bm{A}$ is the adjacency matrix and $\bm{D}$ is the diagonal degree matrix of graph $\mathcal{G}$.
The dimension of the embedding vector of nodes, denoted as $d$, is significant for computational considerations.
In feature propagation, $d$ is the dimension of the embedding vector of nodes, while in label propagation, $d$ is the dimension of the label vector of nodes.

% \minisection{Extension from Top-1 to Top-N Recommendations.}
% To extend the \textsf{MAGUS} system from top-1 recommendations to top-N recommendations, we recursively select a list of $N$ $a_\mathtt{MAGUS}$s as recommendations instead of selecting only one recommendation $a_\mathtt{MAGUS}$, as shown in line~\ref{line:1} in Algorithm~\ref{algo:main}.
% As a result, some modifications in the handling of user feedback are necessary.
% We leave the detailed descriptions and discussions in Appendix~\ref{app:topn}.

\subsection{Connections to Rule-based Algorithms}
Our \textsf{MAGUS} system can be conceptualized as an amalgamation of rule-based and learning-based recommendation methods. 
To elucidate, our approach integrates label propagation, which embodies elements of both rule-based and learning-based strategies.
Concretely, we delineate our framework as a node classification task within the context of a relational graph, aiming to recommend a node to a given user. 
In each $k$-th interaction, our interactive cycle is characterized by an update formula $\bm{Y}^{(k+1)}=\lambda \bm{S}\bm{Y}^{(k)}+(1-\lambda)\bm{Y}^{(0)}$, where $\bm{Y}^{(k)}$ represents the label matrix at the commencement of the $k$-th iteration.
Here, $\bm{S}$ is the symmetric normalized version of the adjacency matrix $\bm{A}$ given by $\bm{S}=\bm{D}^{-\frac{1}{2}}\bm{A}\bm{D}^{-\frac{1}{2}}$, and $\bm{D}$ represents the degree matrix of the graph $\bm{A}$.
Our adjacency matrix $\bm{A}$ is constructed by integrating rules derived from the user's browsing patterns, as described in \emph{Definition~\ref{def:graph}}.
This incorporation of user behaviors into the graph's structure provides a personalized context for the recommendation process.
Additionally, as detailed in Section~\ref{sec:recommendation}, the initial label matrix $\bm{Y}^{(0)}$ is derived from an offline-tuned recommendation model.
And, the weights for the graph edges $\{w_{vv'}|\langle v,v'\rangle \in \mathcal{R}\}$, as discussed in Section~\ref{sec:weight}, also can be computed using an offline-learned recommendation model.

Therefore, our \textsf{MAGUS} system is distinguished by its superior generalization capabilities and its enhanced capacity to manage complex user patterns that may not conform to predefined rules.
This attribute sets it apart from traditional rule-based algorithms, which, while they offer clear interpretability, can sometimes fall short in adapting to the intricacies of user behavior that extend beyond established norms.
In comparison to purely learning-based algorithms, the \textsf{MAGUS} system offers a distinct advantage in terms of interpretability.
This system's design, which incorporates a structured graph with labeled nodes, facilitates a more transparent understanding of its decision-making process.
By examining the graph structure and the labels assigned to individual nodes, one can more readily deduce the rationale behind the system's actions.

We also note that our \textsf{MAGUS} system can ensure the recommendations concentrate on the user feedback within the current session rather than previous user behaviors.
In other words, our \textsf{MAGUS} prioritize the user's current preferences, acknowledging that user interests are subject to change over time. 
For instance, while a user's browsing history may indicate a significant interest in items such as ``iPhone'' items, if the user clearly expresses a preference for other items such as ``MacBook'' items during the current session, the \textsf{MAGUS} system will dynamically adjust its recommendations accordingly.
It will prioritize recommending ``MacBook'' items over ``iPhone'' items.
This is achieved by the design of the mechanisms within \textsf{MAGUS}.
Specifically, according to Eqs.~(\ref{eqn:query}) and (\ref{eqn:norm}), even if a user has shown great interest in some ``iPhone'' items in their previous behaviors, the prediction scores for all ``iPhone'' items are constrained to be smaller than 1.
Furthermore, if a user clearly expresses a preference for ``MacBook'' items, as per Eq.~(\ref{eqn:user}), the prediction scores for ``MacBook'' items are set to 1.
If all edge weights are set to 1 for $\mathcal{R}^+$ (following Eq.~(\ref{eqn:weight})), items related to ``MacBook'' items would receive prediction scores greater than 1 according to Eq.~(\ref{eqn:update}).
In this case, it is unlikely for our \textsf{MAGUS} system to respond with an ``iPhone'' items when a user clearly prefers ``MacBook''.

\section{Experiment Configurations}
\subsection{Dataset Description}
\label{app:dataset}
We conducted extensive experiments on 3 industrial real-world e-commerce datasets, whose statistics are summarized as follows. 
\begin{itemize}[topsep = 3pt,leftmargin =5pt]
\item \textbf{Amazon}\footnote{\url{https://jmcauley.ucsd.edu/data/amazon/}}
is a dataset introduced by collected from Amazon, an online e-commerce application from May 1996 to July 2014.
There are 1,114,563 reviews of 133,960 users and 431,827 items with an average sequence length of 33 and 6 feature fields. 
\item \textbf{Alipay}\footnote{\url{https://tianchi.aliyun.com/dataset/dataDetail?dataId=53}}
is a dataset collected by Alipay, an online payment application from July 2015 to November 2015.
There are 35,179,371 interactions of 498,308 users and 2,200,191 items with an average sequence length of 70 and 6 feature fields.
\item \textbf{Tmall}\footnote{\url{https://tianchi.aliyun.com/dataset/dataDetail?dataId=42}}
is a dataset consisting of 54,925,331 interactions of 424,170 users and 1,090,390 items.
These sequential histories are collected by Tmall e-commerce platform from May 2015 to November 2015 with an average sequence length of 129 and 9 feature fields. 
\end{itemize}

\subsection{Description of Data Pre-processing}
\label{app:preprocess}
For each dataset, we begin with organizing users and items into a set of users $\mathcal{U}$ and a set of items $\mathcal{V}$.
Also, we can get access to user-item interaction histories to build: $\mathcal{H}^+_u$, a set of items receiving positive feedback from user $u$; and $\mathcal{H}^-_u$, a set of items receiving negative feedback from user $u$, for all user $u\in\mathcal{U}$.
As outlined in \emph{Definition~\ref{def:task}}, we need a query set $\mathcal{Q}$.
In the following, we take the Amazon dataset as a concrete example of the generation of $\mathcal{Q}$.
Initially, we create a word set $\mathcal{W}$ by extracting terms from the categorical features of items within $\mathcal{V}$.
For instance, if there is an item with brand \texttt{Coxlures} and belonging to categories such as \texttt{Sports} and \texttt{Dance}, then we extract words \texttt{coxlures}, \texttt{sports}, and \texttt{dance}.
We proceed by enumerating all the items in $\mathcal{V}$.
During this process, we form combinations of words and individual words as the nodes of our relational graph.
For example, if we have an \texttt{Item} \texttt{A} with a combination like \texttt{coxlures}, \texttt{sport} \texttt{and} \texttt{dance}, then we have nodes as individual words such as \texttt{coxlures}, \texttt{sport}, and \texttt{dance}, and nodes as combinations of words such as \texttt{sport} \texttt{and} \texttt{dance}.
All these nodes, except for those representing items, can be considered as queries (i.e., $\mathcal{Q}$). 
They serve as potential query elements within our system.
Also, once we have established the nodes representing queries and items, we can proceed to construct our relational graph $\mathcal{G}$ (as defined in \emph{Definition}~\ref{def:graph}) correspondingly.

\subsection{Description of Relational Graph Construction}
\label{app:graph}
We also emphasize the versatility of our relational graph $\mathcal{G}$, which can be constructed across various domains. 
For example, when dealing with items like movies or songs, we can create these graphs using key features extracted from raw data. For movies, features such as directors and genres can be extracted, while for songs, features like singers and genres can be utilized. 
These features form nodes $\mathcal{V}$ within the graph, representing combinations of attributes (e.g., \texttt{director} \texttt{A} \texttt{+} \texttt{romantic}, \texttt{director} \texttt{B} \texttt{+} \texttt{fiction}). 
We define mutual improvement relationship between nodes $\mathcal{R}^+$ like \texttt{director} \texttt{A} and \texttt{director} \texttt{A} + \texttt{romantic}, as users inclined towards the romantic genre may favor movies directed by \texttt{director} \texttt{A} in that genre. 
Conversely, we define mutual inhibition relationship $\mathcal{R}^-$ between nodes like romantic and fiction, reflecting the understanding that users preferring fiction may not favor movies categorized under the romantic genre. 
Additionally, we define a mutual independence relationship between nodes $\mathcal{R}^{\bot}$ between nodes like nodes representing directors and nodes representing genres, because there are no explicit relations between these two types of nodes, and the implicit relations are modeled by introducing nodes like \texttt{director} \texttt{A} + \texttt{romantic}.
These relationships capture underlying interdependencies and refine the system's recommendations. 
Our primary focus lies in strategically leveraging relational graphs in multiple-round settings, while the extraction of relevant information, such as keywords, from item descriptions aligns with established tokenization practices in the field \citep{mielke2021between}, which is out of the scope of this paper.

\subsection{Descriptions of Adjustments of Baselines}
\label{app:baseline}
\begin{figure}[h]
	\centering
	\includegraphics[width=0.8\linewidth]{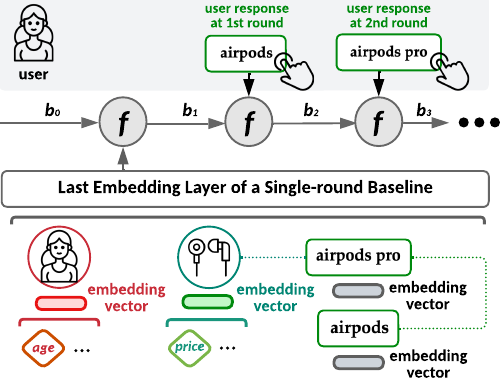}
	\vspace{-3mm}
	\caption{An illustrated example of extending a single-round baseline method to our multiple-round setting in the masonry layout, where we introduce a recurrent neural network to facilitate the flow of information across rounds.
	}
	\label{fig:baseline}
	\vspace{-2mm}
\end{figure}

Consider that our task of multiple-round recommendations supporting queries and items (as introduced in \emph{Definition}~\ref{def:task}) is novel, and there are no existing approaches tailored to this specific task.
To adapt the baseline methods to our context, we introduce two main adjustments.
The first adjustment involves expanding their action space from just items (i.e., $\mathcal{V}$) to encompass both items and queries (i.e., $\mathcal{A}=\mathcal{V}\cup\mathcal{Q}$).
The second adjustment enables these methods to utilize user responses during multiple-round interactions.
We introduce a recurrent neural network architecture, outlined in Figure~\ref{fig:baseline}, to facilitate this process. 
Initially, we input the final embedding layer (i.e., the representation of the final prediction) of a single-round baseline method into a unit, enabling information from each round to be transmitted to subsequent rounds. 
Inspired by the label trick proposed in \citep{jin2022learn}, we devise representation vectors to capture user behaviors, such as clicks. 
In the case of binary user responses, we utilize two vectors to represent positive and negative user responses. These representation vectors for user feedback are concatenated with user and item features, forming the input of the model.

\subsection{Implementation Details}
\label{app:code}
The learning rate is decreased from the initial value $1\times 10^{-2}$ to $1\times 10^{-5}$ during the training process.
The dimensions of all the embedding vectors including users and words (and queries and items) are set as 64.
The batch size is set as 1000.
The weight for L2 regularization term is $4\times 10^{-4}$.
The dropout rate is set as 0.5.
We assign the length of recommendation lists (denoted as $N$) as 3. 
The maximum number of rounds (denoted as $K_\mathtt{MAX}$) is set in the evaluation metrics (see Section~\ref{sec:exp} for details). 

\section{Simulation Design}
\subsection{Description of Conversational Agent}
\label{app:agent}
Our simulator introduces a conversational agent built upon the recommendation model to interact with a human user.
The following summarizes the interaction principles.

\begin{definition}
[\emph{\textbf{\textsf{Conversational Agent and Human User Interactions}}}]
Our conversational agent is designed to act in the following 2 ways following Eq.~(\ref{eqn:recommend}).
\begin{itemize}[topsep = 0pt,leftmargin =15pt]
\item Query an item $v\in\mathcal{V}$, when the corresponding node receives the highest score.
\item Query a query $q\in\mathcal{Q}$, when the corresponding node (representing a specific word combination) receives the highest score.
\end{itemize}
The human user is supposed to respond in the following ways.
\begin{itemize}[topsep = 0pt,leftmargin =15pt]
\item Upon querying item $v$, the user should respond with \texttt{Yes} if $v$ is one of target items of users, or \texttt{No} otherwise. 
In the ``\texttt{Yes}'' case, the session is successfully concluded, because $v$ is one of the target items (i.e., the user has identified an item that meets their satisfaction criteria).
Conversely, in the \texttt{No} case, the scores assigned to each node are subjected to an update process. 
\item Upon querying query $q$, the user should respond with \texttt{Yes} if $q$ is one of the elements within the word combination of the target items, or \texttt{No} otherwise.
The scores assigned to each node would be updated accordingly.
\end{itemize}
\end{definition}
We note that in the following definition when a session encompasses multiple target items, the conversational agent's requirement is limited to identifying a single target item that meets the user's criteria.
One underlying assumption under this setting is that users always have a clear picture of their interests, which may not hold in practice, as users often have multiple interests and some of them would be ambiguous or not clearly defined.
For instance, as depicted in Figure~\ref{fig:overview}, a user may simultaneously express a preference for both \texttt{Item} \texttt{A} and \texttt{Item} \texttt{B}.
If the conversational agent inquire about \texttt{airpods}, the user agent is expected to affirm with a ``\texttt{Yes}'', thereby directing the recommender system to identify \texttt{Item} \texttt{B} as a potential match.
However, in practical scenarios, the user might also favor \texttt{Item} \texttt{A} concurrently, given that the user's preferences are not mutually exclusive.

Therefore, we introduce another setting that permits the user to respond with \texttt{Not} \texttt{Care} when faced with a query from the conversational agent that pertains to attributes or features present in multiple preferred items.
This response option acknowledges the ambiguity of user preferences and allows for a more nuanced interaction with the recommender system.
The corresponding results under this setting are available in Appendix~\ref{app:robustness}. 

\subsection{Empowering MAGUS with Large Language Models}
\label{app:llm}
As depicted in Figure~\ref{fig:motivation}, our \textsf{MAGUS} system integrates both queries and items in its output layout.
However, a potential concern with this design is the occasional generation of queries that may seem artificial, potentially diverging from user-friendly formulations. 
To mitigate this challenge, a feasible strategy entails harnessing the capabilities of large language models while employing targeted prompts to steer the generation of queries toward user-friendly, natural language formulations.

Initially, if the expectation is for the final queries to consist of combinations of keywords, the following prompt can be utilized:

\noindent
\emph{prompt = f"""}

\noindent
\emph{You will be provided with text delimited by triple quotes.}

\noindent
\emph{If words in the text can be keywords of a query, then you just return it. Otherwise, you should re-organize these words. You should just return the original or re-organized words, not sentences.}

\noindent
\emph{```{text}'''"""}

Secondly, if we anticipate the final queries to be in the form of completed sentences, we can utilize the following prompt.

\noindent
\emph{prompt = f"""}

\noindent
\emph{You will be provided with text delimited by triple quotes.}

\noindent
\emph{The input text can be regarded as a class of items. You should generate a sentence to ask whether the user likes this class of items or not. You should be gentle.}

\noindent 
\emph{```{text}'''"""}

Next, we proceed to assess the performance of the aforementioned methods as follows.
The first prompt is tailored to extract keywords from the input text while eliminating redundancy.
For instance, if the input text is ``\emph{blue clothes shirt}'', the output would be ``\emph{blue shirt}'' with ``\emph{clothes}'' considered redundant due to the presence of ``\emph{shirt}'' in the query.
The second prompt is designed to generate a sentence or conversation based on the query.
For example, if the input text is ``\emph{blue shirt}'', the output would be ``\emph{Do you like blue shirts?}''
This output constitutes a complete sentence suitable for use in a conversation.

We also note that LLM can encode some latent features from the conversations.
For example, we can add the following statement in the 
input prompt: ``\emph{If you can infer the user has a low budget, please make recommendations with low prices}''.
In response to the user mentioning ``\emph{I am a student}'', the system would subsequently provide queries or items characterized by low prices.

Considering that the above process only requires a large language model API, it is very practical to integrate a large language model into our \textsf{MAGUS} system as later discussed in Appendix~\ref{app:arch}.

\begin{table}[t]
	\centering
	\caption{Results comparison of joint recommendations of both queries and items in terms of RA@3, SA@3, and SA@5 under the user agent configuration introduced in Appendix~\ref{app:agent}.
	* indicates $p < 0.001$ in significance tests compared to the best baseline.
	}
	\vspace{-3mm}
	\resizebox{1\linewidth}{!}{
		\begin{tabular}{@{\extracolsep{4pt}}|c|ccc|ccc|}
		\toprule
			\multirow{2}{*}{Methods} & \multicolumn{3}{c}{Amazon} |& \multicolumn{3}{c}{Alipay} |\\
			\cmidrule{2-4}
			\cmidrule{5-7}
			{} & RA@3 & SA@3 & SA@5 & RA@3 & SA@3 & SA@5 \\
                \midrule
			FM & 
			0.685 & 0.602 & 0.658 & 
			0.720 & 0.700 & 0.762 \\
                \midrule
                FM+CRM & 
                0.702 & 0.610 & 0.668 & 
			0.735 & 0.725 & 0.776 \\
                \midrule
                FM+ME & 
                0.701 & 0.608 & 0.659 & 
			0.742 & 0.729 & 0.780 \\
                \midrule
                FM+EAR & 
                0.711 & 0.615 & 0.674 & 
			0.740 & 0.728 & 0.786 \\
                \midrule
			FM+\textsf{MAGUS} &
			\textbf{0.725}$^{*}$ & \textbf{0.625}$^{*}$ & \textbf{0.682}$^{*}$ & 
			\textbf{0.759}$^{*}$ & \textbf{0.750}$^{*}$ & \textbf{0.803}$^{*}$ \\
                \midrule
			DeepFM & 
			0.702 & 0.612 & 0.706 & 
			0.736 & 0.698 & 0.788 \\
                \midrule
                DeepFM+CRM & 
                0.716 & 0.624 & 0.719 & 
			0.750 & 0.713 & 0.805 \\
                \midrule
                DeepFM+ME & 
                0.709 & 0.615 & 0.704 & 
			0.740 & 0.704 & 0.791 \\
                \midrule
                DeepFM+EAR & 
                0.742 & 0.646 & 0.746 & 
			0.771 & 0.735 & 0.824 \\
                \midrule
			DeepFM+\textsf{MAGUS} &
			\textbf{0.759}$^{*}$ & \textbf{0.655}$^{*}$ & \textbf{0.760}$^{*}$ & 
			\textbf{0.785}$^{*}$ & \textbf{0.749}$^{*}$ & \textbf{0.840}$^{*}$ \\  
			\bottomrule
		\end{tabular}
	}
	\label{tab:appres}
	\vspace{-2mm}
\end{table}

\begin{figure}[h]
	\centering
	\vspace{-1mm}
	\includegraphics[width=0.82\linewidth]{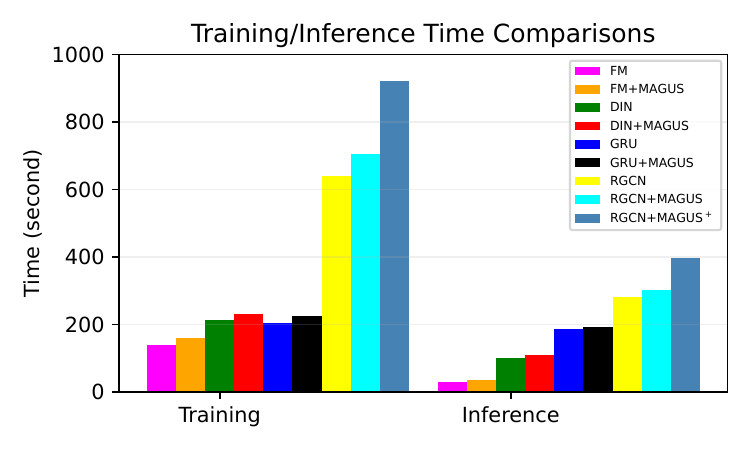}
	\vspace{-6mm}
	\caption{Comparisons of training and inference time of \textsf{MAGUS} and \textsf{MAGUS}$^+$ against several baselines on Tmall dataset.}
	\label{fig:time}
	\vspace{-2mm}
\end{figure}

\section{Additional Experiments}
\subsection{Performance Comparisons with Ambiguous User Feedback}
\label{app:robustness}
As outlined in Appendix~\ref{app:agent}, the user agent, used in Section~\ref{sec:exp}, is designed to respond with ``\texttt{Yes}'' to the conversational agent if the query matches any of the designated target items.
However, this assumption may not always be valid in practice.
Consequently, an alternative configuration is presented in Appendix~\ref{app:agent} to account for the variability and potential ambiguity inherent in user feedback.

Table~\ref{tab:appres} reports the corresponding results on Amazon and Alipay datasets.
The table demonstrates that our \textsf{MAGUS} system consistently outperforms the baseline methods, even when considering the ambiguity of user feedback, thereby showcasing the robustness of our system.
Furthermore, a comparison between Table~\ref{tab:appres} and Table~\ref{tab:res} reveals that ambiguity significantly impairs performance, particularly for X and X+ME.
One possible explanation is that X, which incorporates user feedback as input, inevitably introduces more noise, and X+ME, which emphasizes querying over item recommendation, is more susceptible to user ambiguity, as ambiguity frequently arises during the querying process.

\subsection{Complexity Study}
\label{app:time}
We conducted an investigation into the time complexity of \textsf{MAGUS} and \textsf{MAGUS}$^+$ based on the FM, GRU, DIN, and RGCN, recommendation bases on Tmall dataset.
We report the training and inference times for one round of the whole data in Figure~\ref{fig:time}.
In the training phase, \textsf{MAGUS} showed almost no significant difference in terms of computational time when compared to the baseline methods. This is because the computations introduced by \textsf{MAGUS} mainly involve building the relational graph, which is relatively efficient.
In the inference phase, the computational costs of both \textsf{MAGUS} and \textsf{MAGUS}$^+$ are incurred by the label propagation algorithm. 
The use of \textsf{MAGUS}$^+$ introduces some additional computational overhead compared to \textsf{MAGUS}, as it involves running the one-round feature propagation on the graph.
% Overall, the results indicate that the computational overhead introduced by \textsf{MAGUS} and \textsf{MAGUS}$^+$ in the inference phase is manageable and does not significantly impact the efficiency of the recommendation system.

\subsection{Case Study}
\label{app:case}
\begin{figure}[h]
	\centering
% 	\vspace{-2mm}
	\includegraphics[width=1.0\linewidth]{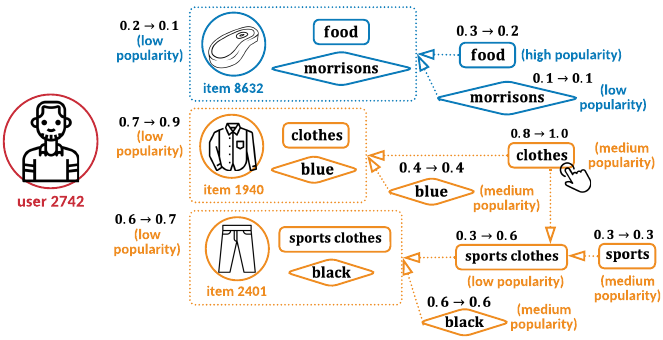}
	\vspace{-6mm}
	\caption{
	    An illustrated case of \textsf{MAGUS} providing one recommendation to user \texttt{2742} at each round.
	The number depicted over each node is its score regarding the user. 
 The popularity level of each node is computed according to its frequency in the user's browsing log.
	}
	\label{fig:case}
	\vspace{-2mm}
\end{figure}

In order to further demonstrate the superiority of \textsf{MAGUS}, we conduct a case study on Amazon dataset.
Figure~\ref{fig:case} illustrates the process of \textsf{MAGUS}.
\textsf{MAGUS} first initializes scores for nodes in the relational graph.
These nodes represent individual words and combinations of words, serving as queries and items.
After receiving responses from a specific user, \textsf{MAGUS} updates the scores accordingly.
This dynamic process allows \textsf{MAGUS} to adapt to the user's preferences and provide personalized recommendations.

We argue that this practical case holds significance because it contrasts a simple and commonly used method \citep{bar2011context} that always recommends highly popular queries or items to users (denoted as MPS in our experiment).
For clarity, we have also incorporated a visualization of each node's popularity level in the figure.
In the specific instance, \textsf{MAGUS} recommends \texttt{clothes} as a query in the first round and item \texttt{1940} in the second round, while the popularity-based method would suggest \texttt{food} in the first round and \texttt{clothes} in the second round.

This comparison effectively underscores a key point: popularity-based methods tend to overlook the variations in user preferences and the dynamic nature of these preferences across different rounds.
This drawback resembles the well-known popularity bias \citep{chen2023bias} seen in item recommendation tasks, where popular items are favored in recommendations without due consideration of individual user preferences.

\begin{figure}[h]
	\centering
	\includegraphics[width=1.00\linewidth]{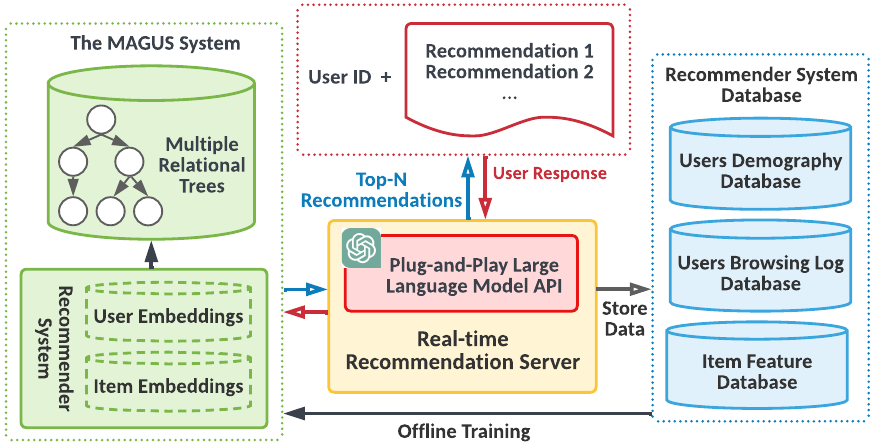}
	\vspace{-6.5mm}
	\caption{Online deployment of the \textsf{MAGUS} system can be seamlessly integrated into existing real-time recommendation servers. The diagram below shows the workflow, where solid arrows represent the offline training phase, and hollow arrows represent the online update phase. 
    Most of the components required for the \textsf{MAGUS} system are already present in existing recommender systems.
    The main additions are the multiple relational trees, which store the relational graph used by \textsf{MAGUS}.
    We also incorporate large language models in a plug-and-play manner to decorate the artificial queries generated by \textsf{MAGUS}.}
	\label{fig:deploy}
	\vspace{-2mm}
\end{figure}

\section{Deployment Feasibility}
\subsection{System Architecture for Deployment}
\label{app:arch}
Here, we discuss the feasibility of the industrial deployment of our \textsf{MAGUS} system.
Fortunately, integrating the \textsf{MAGUS} system into an existing recommendation platform should not pose a substantial workload.
The primary modifications introduced by \textsf{MAGUS} involve the organization of the relational graph and its continuous updating based on user responses.
The majority of the prediction model pipeline remains unchanged.

Integrating the \textsf{MAGUS} system into an existing recommendation platform should not introduce a significant computational burden, since the primary modifications only involve organizing the relational graph and updating the graph by user responses.

To optimize the computation costs associated with using the graph multiple times, we propose a strategy for simplification.
This involves creating multiple relational trees, as illustrated in Figure~\ref{fig:deploy}.
Each of these trees originates from a root node, which typically represents an item, and then branches out to nodes representing individual words.
As an example, consider Figure~\ref{fig:overview}(a).
In this case, one of these trees extends from a node representing \texttt{Item} \texttt{A} and branches to nodes that represent individual words like \texttt{apple} and \texttt{grey}.
The primary benefit of this approach is the ability to parallelize computations and focus on localized operations within each tree.
This parallelization can lead to significantly improved efficiency when working with the graph in a multiple-round recommender system.

\subsection{Extension of \textsf{MAGUS} from Top-1 to Top-N Recommendations}
\label{app:topn}
To extend \textsf{MAGUS} from top-1 recommendations to top-N recommendations, we only need to make the following modifications to Algorithm~\ref{algo:main}. 

First, instead of selecting only one recommendation $a_\mathtt{MAGUS}$ in line~\ref{line:1}, we now recursively select a list of $N$ $a_\mathtt{MAGUS}$s as recommendations.
Any one of these $a_\mathtt{MAGUS}$s can represent either a query or an item, and they are chosen based on the top-$N$ predicted scores from the recommender system $\psi_\mathtt{RE}(\cdot)$.

Second, the user's response and the subsequent updates differ from lines~\ref{line:2} and \ref{line:3}. 
In the case of binary user response (namely, the user response is either positive or negative) to a recommendation, the response to top-N recommendations fails to two categories.
(i) If the user clicks on one of the $N$ $a_\mathtt{MAGUS}$s, denoted as $a^*_\mathtt{MAGUS}$, we set the score of the node representing $a_\mathtt{MAGUS}$ to 1, and the scores of the nodes representing the remaining $N-1$ $a_\mathtt{MAGUS}$ are set to 0.
This adjustment is accomplished by modifying Eq.~(\ref{eqn:user}) accordingly.
(ii) If the user does not click on any of the $N$ $a_\mathtt{MAGUS}$s, we assign scores of 0 to the nodes representing all $N$ $a_\mathtt{MAGUS}$s.
For any category of user responses, we subsequently apply Eq.~(\ref{eqn:update}) to propagate these updated scores to the other nodes in the graph.

These modifications enable the \textsf{MAGUS} system to provide top-N recommendations and handle user responses accordingly.

\subsection{Deployment Feasibility for Exploratory Search}
\label{app:explore}

\begin{figure}[h]
	\centering
% 	\vspace{-2mm}
	\includegraphics[width=0.95\linewidth]{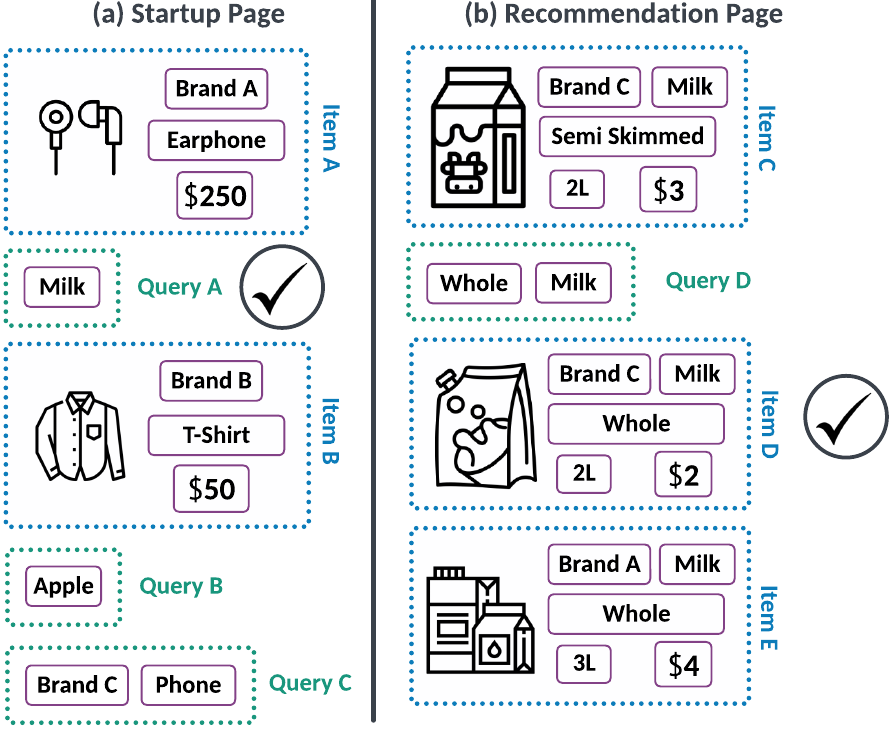}
	\vspace{-2mm}
	\caption{
	    An illustrated example of exploratory search is where users often have a vague idea of the items they seek.
     \textsf{MAGUS} provides a selection of recommended items alongside suggested queries, allowing users to engage in navigating and comparing items prior to refining their search queries.
	}
	\label{fig:explore}
	\vspace{-2mm}
\end{figure}

\label{app:explore}
Here, we extend the example in Figure~\ref{fig:motivation}(b) to further illustrate how our \textsf{MAGUS} system would enable the exploratory search where users often have a vague idea of the items they seek. 
As introduced in Figure~\ref{fig:explore}, a user tends to purchase milk but lacks specific criteria such as
the preferred brand or price range to give a well-formed query. 
In such an instance, one viable solution is to provide a selection of recommended milk items alongside suggested queries, allowing users
to engage in navigating and comparing items prior to refining their
search queries.
Therefore, the user initially chooses to search for the query \texttt{Milk} from the startup page, comparing it with other presented queries and items, as depicted in Figure~\ref{fig:explore}(a).
Subsequently, \textsf{MAGUS} provides the user with multiple queries and items, enabling her to compare suggested queries (at an abstract level) or recommended items (at a specific level) to determine whether to search for a more specific query or directly select an item.
As a result, the user chooses item D.
From the above example, we can see that achieving this requires jointly considering queries and items; however, almost all the deployments
of either query recommendations or item recommendations have predominantly occurred in isolation from one another.

% We provide the empirical results of time complexity, along with discussions on deployment feasibility and exploratory search, in the supplementary material.

\end{document}